\documentclass[aps,prd,10pt,notitlepage,nofootinbib,superscriptaddress,showkeys,showpacs]{revtex4-1}

\usepackage{amsmath,amssymb}
\usepackage[english]{babel}
\usepackage{graphicx,color}
\usepackage{xspace}
\usepackage{graphicx}
\usepackage{pifont,dsfont}
\usepackage{marvosym}
\usepackage{slashed}
\usepackage{subfigure}

\newcommand{\be}{\begin{equation}}
\newcommand{\ee}{\end{equation}}
\newcommand{\bqa}{\begin{eqnarray}}
\newcommand{\eqa}{\end{eqnarray}}
\newcommand{\bea}{\begin{eqnarray}}
\newcommand{\eea}{\end{eqnarray}}

\newcommand{\NN}{{\mathbb  N}}

\newtheorem{definition}{Definition}
\newtheorem{theorem}{Theorem}

\newtheorem{proposition}{Proposition}

\newcommand{\cM}{{\cal M}}

\DeclareMathOperator{\tr}{tr}

\DeclareMathOperator{\diff}{d\!}

\begin{document}

\title{\Large \bf Tensor models from the viewpoint of matrix models: the case of loop models on random surfaces}

\author{{\bf Valentin Bonzom}}\email{bonzom@lipn.univ-paris13.fr}
\affiliation{LIPN, UMR CNRS 7030, Institut Galil\'ee, Universit\'e Paris 13, Sorbonne Paris Cit\'e,
99, avenue Jean-Baptiste Cl\'ement, 93430 Villetaneuse, France, EU}
\author{{\bf Fr\'ed\'eric Combes}}\email{frederic.combes@ens-lyon.fr}
\affiliation{Perimeter Institute for Theoretical Physics, 31 Caroline St. N, ON N2L 2Y5, Waterloo, Canada}

\date{\small\today}

\begin{abstract}
\noindent We study a connection between random tensors and random matrices through $U(\tau)$ matrix models which generate fully packed, oriented loops on random surfaces. The latter are found to be in bijection with a set of regular edge-colored graphs typically found in tensor models. It is shown that the expansion in the number of loops is organized like the $1/N$ expansion of rank-three tensor models. Recent results on tensor models are reviewed and applied in this context. For example, configurations which maximize the number of loops are precisely the melonic graphs of tensor models and a scaling limit which projects onto the melonic sector is found. We also reinterpret the double scaling limit of tensor models from the point of view of loops on random surfaces. This approach is eventually generalized to higher-rank tensor models, which generate loops with fugacity $\tau$ on triangulations in dimension $d-1$. 



\end{abstract}

\medskip


\maketitle

\section*{Introduction}

Matrix models have been very useful to generate and study random geometries in two dimensions. At large matrix size $N$, the $1/N$ expansion is a topological expansion, labeled by the genus of the random discrete surfaces. In the large $N$ limit, only planar maps on the sphere survive. These maps encode discrete geometries of fluctuating surfaces, making them very important in physics. A famous application is two-dimensional gravity coupled to conformal matter (central charge $c<1$) \cite{mm-review-difrancesco}.

Tensor models allow to extend those ideas to random geometries with more than two dimensions \cite{ambjorn-3d-tensors, sasakura-tensors, gross-tensors}. Their Feynman expansion is a sum over discretized (pseudo-)manifolds in dimension $d$ and it possesses a $1/N$ expansion \cite{Gur4, uncoloring}. A continuum limit exists, first found in \cite{critical-colored}, which can be coupled to (non-unitary) critical matter \cite{harold-hard-dimers, multicritical-dimers}, leading to different universality classes.

The progress obtained in the past few years on tensor models are due to the discovery that tensor models with a $U(N)^d$ symmetry naturally generate regular, edge-colored graphs (dual to triangulations of pseudo-manifolds) \cite{uncoloring}. Those graphs, in contrast with the stranded graphs initially considered in \cite{ambjorn-3d-tensors}, are amenable to analytical investigation. A combinatorial classification has been recently obtained, \cite{GurauSchaeffer}. In the same time, tensor models with quartic interactions have been re-formulated as matrix models, \cite{DSQuartic, GenericQuartic}. Both approaches have led to a double-scaling limit and more generally to a good understanding of the singularities at fixed order in the $1/N$ expansion. The double scaling limit has been extended to models beyond the quartic interactions in \cite{DSSD} using a typical tool of matrix models, the loop equations.

It thus appears that matrix model techniques can be useful in tensor models. Formulating tensor models as matrix models also opens the possibility of using the combinatorial techniques (or even maybe already existing results) on maps. However a precise study of the relationships between tensor and matrix models has not appeared yet. This is the program we start in the present article.

It was not obvious at first that matrix models techniques would be of any use. In particular, diagonalization and eigenvalues (together with the saddle point method or orthogonal polynomials) are among the most effective tools in random matrix models and they are not available for random tensors. Also the fact that $U(N)^d$-invariant random tensors become Gaussian at large $N$ \cite{universality}, and are thus very different from large $N$ matrix models, tends to establishing a clear distinction between matrices and tensors.

However those arguments are no longer relevant thanks to the intermediate field method which turns quartic models into multi-matrix models. In addition, there are two simple ideas which establish a direct connection between matrices and tensors, which we present and exploit in this article. They enable to understand the position of tensor models with respect to matrix models. Following those two ideas one after the other, we offer a novel presentation of random tensor models, in which results from tensor models are applied to matrix models and the other way around.

The first part \ref{sec:loops} is based on the observation that a collection of matrices $M_1,\dotsc, M_\tau$ may be packaged into a tensor of rank three and size $N\times N\times \tau$, whose first two indices are matrix indices while the third one is the label of the matrix. When the joint probability distribution on the matrices is of the form $e^{-V}$ for a polynomial potential $V$ that is $U(\tau)$-invariant, then we have a tensor model in disguise. We therefore introduce a family of $U(\tau)$ models which is shown to generate random surfaces dressed with configurations of oriented loops. We describe the bijection between the observables and the Feynman graphs of those $U(\tau)$ models and their corresponding tensor models.

All known $1/N$ expansions in tensor models rely on the \emph{degree} of the Feynman graphs dual to the triangulations. It was originally introduced in \cite{Gur3} to exhibit a $1/N$ expansion for tensor models for the first time. The degree was defined as a sum of genera of ribbon sub-graphs which are generated by matrix models embedded in the tensor theory \cite{JacketsMatrixModels}. It controls the balance between the number of faces and the number of vertices and reduces to the genus in two dimensions. The dominant triangulations of tensor models at large $N$ are those with vanishing degree and are known as \emph{melonic} triangulations \cite{critical-colored, uncoloring}, which have a specific, highly curved, geometry. They have been recently matched to random branched polymers \cite{melons-BP}, meaning that the continuous geometry is that of the continuous random tree. Melonic graphs are the ones which maximize the number of faces at fixed number of vertices \cite{critical-colored}.

In the $U(\tau)$ models, it turns out that the number of loops at fixed genus of the random surfaces, fixed numbers of edges and vertices, is counted by the degree of the 4-colored graph representative of the Feynman graph. This provides a new combinatorial interpretation of the degree. In particular, melonic graphs are those which maximize the number of loops. It also makes clear how the large $N$, melonic behavior of tensor models arise from a matrix model when $\tau\to\infty$. We then apply the classification of edge-colored graphs from \cite{GurauSchaeffer} to the quartic $U(\tau)$ model to get a classification of its loop configurations. Finally, the double scaling limit of tensor models is found to resum consistently the most critical loop configurations.

The intermediate field transformation is also performed on the quartic $U(\tau)$ models, leading to a two-matrix model. To our knowledge, this two-matrix model has never been studied in the matrix model literature and we do not even know its large $N$ free energy. It generates graphs formed by two maps glued together at their vertices (at least at one of them for the whole graph to be connected). Those graphs, also called \emph{nodal surfaces} do have already appeared in the literature \cite{EynardBookMulticut} and they may be well suited for a combinatorial analysis.

In a companion paper \cite{AngularIntegralsGaussian}, another simple relationship between matrix and tensor models is studied. It relies on re-packaging the set of $d$ indices into two disjoint sets which are interpreted as indices of range $N^p$ and $N^{d-p}$ so that $T$ becomes a (typically rectangular) matrix. The singular value decomposition then enables to perform partial integration over the angular degrees of freedom. The results in a notion of effective observables which actually allows the calculation of new expectations in the Gaussian distribution.

In addition to exploring relationships between random tensors and matrices, those approaches clarify the difficulties faced by random tensor theory in the light of familiar matrix models. We also hope that it sets a frame in which those challenges may be dealt with.

Finally, the section \ref{sec:interpolating} in appendix investigates the possibility of interpolating matrix and tensor models, a question often asked, or more generally interpolating various tensor models. We use for instance a tensor of size $N\times N \times \dotsb \times N^\beta$ where $\beta$ runs in $[0,1]$. This completes the analysis of \cite{new1/N} of tensor models with distinct index ranges. It is found that there are only two large $N$ behaviors in those models, $\beta=0$ and $\beta\in (0,1]$. The reason is that for $\beta=0$ we have a tensor model of rank $d-1$ but as soon as $\beta>0$ each face of colors $(0,d)$ contributes to the large $N$ scaling. However, the $1/N$ corrections are typically found to depend on $\beta$, but we do not know if this affects the continuum limits.

\section{The degree expansion in completely packed loop models on random surfaces} \label{sec:loops}


\subsection{Loop model on random surfaces}

\begin{figure}
	\center
		\subfigure[Interpretation as crossing loops on the ribbon vertex generated by the term $\tr A_a A_b^\dagger A_b A_c^\dagger A_d A_a^\dagger A_c A_d^\dagger$]{\makebox[6cm]{\includegraphics[width=4cm]{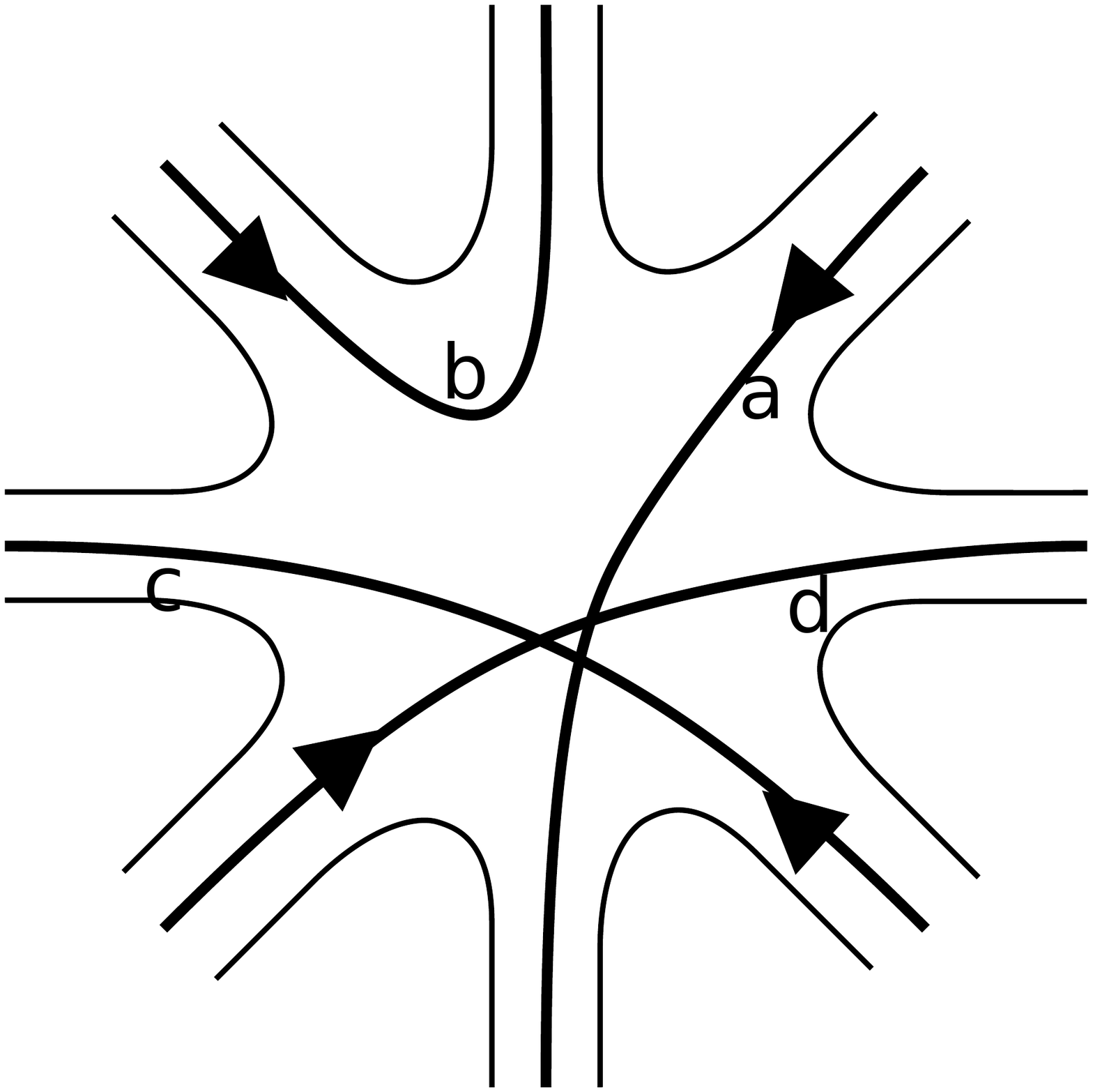}}}
	\hspace{1cm}
		\subfigure[Interpretation as non-crossing loops of the term $\tr A_a A_b^\dagger A_b A_a^\dagger A_c A_c^\dagger A_d A_d^\dagger$\label{subfig:NonCrossingLoop}]{\makebox[6cm]{\includegraphics[width=4cm]{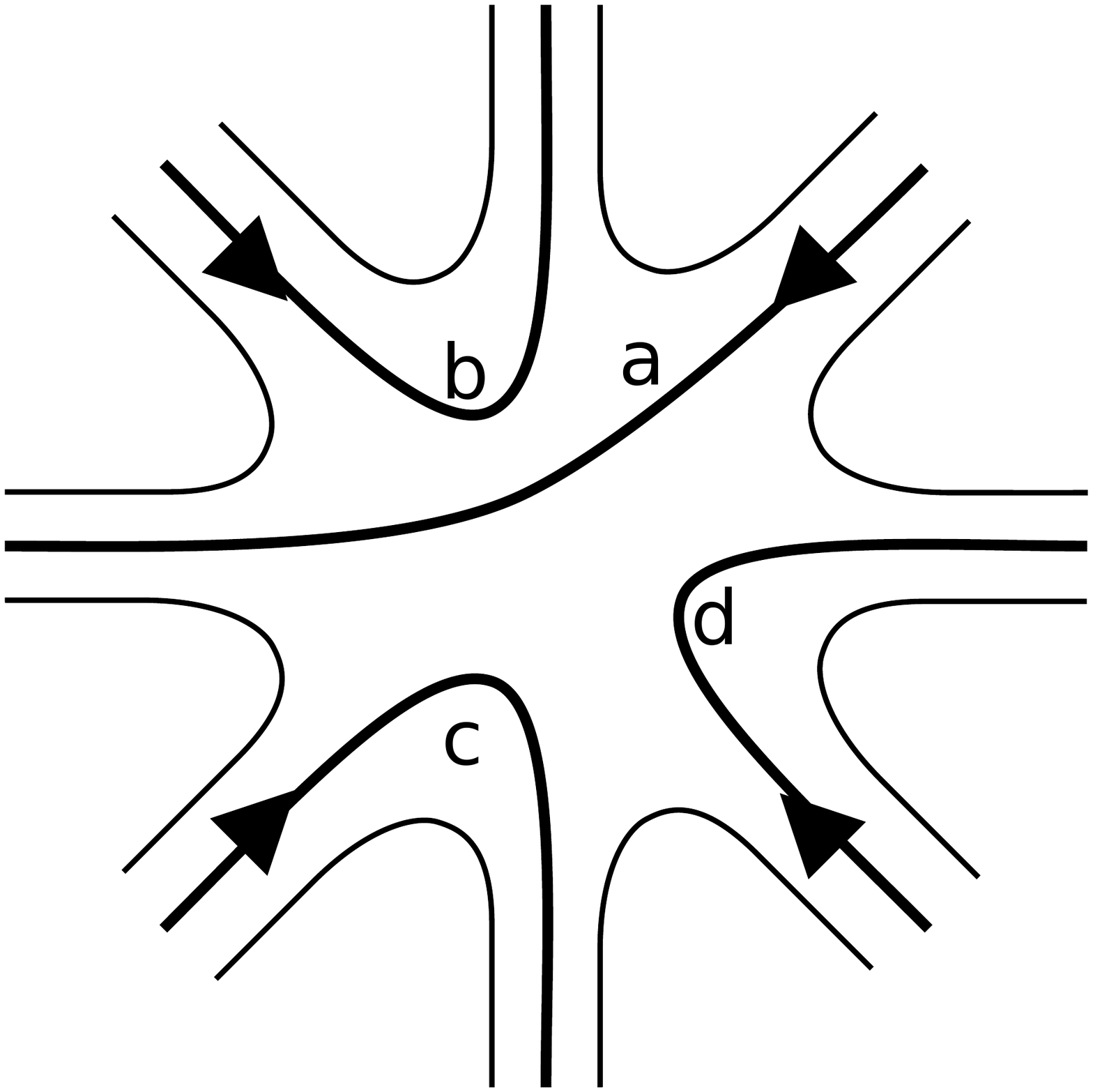}}}
	\caption{Interpretation in terms of loops on ribbon vertices of the labeled matrix model. The loops are naturally oriented from $A_\alpha$ towards $A_\alpha^\dagger$. \label{fig:InterpretationAsLoops}}
\end{figure}

Matrix models are known to generate discretized random 2D surfaces. Each term of the action has the form $\tr (AA^\dagger)^n$, where $A$ is a complex matrix, and creates ribbon vertices of degree $2n$. A matrix model generates random surfaces through the Wick theorem which connects these ribbon vertices together via ribbon lines. Following the recipe of \cite{difrancesco-folding}, the random surfaces can be decorated with oriented loops in the following way: let $\{A_i, A_i^\dagger, i = 1, \dotsc, \tau\}$ be a set of decorated matrices, and rewrite the terms $\tr (AA^\dagger)^n$ with various matrix labelings. We allow terms of the form
\begin{equation}
\label{eq:TraceInvariant}
	V_{n,\sigma}(\{A_i,A_i^\dagger\}) = \sum_{\substack{\alpha_1, \dotsc\alpha_n\\\beta_1,..,\beta_n}} \tr\left( A_{\alpha_1}A^\dagger_{\beta_1} A_{\alpha_2} A^\dagger_{\beta_2}\,\dotsm\,A_{\alpha_n}A^\dagger_{\beta_n}\right) \prod_{k=1}^n \delta_{\alpha_k,\beta_{\sigma(k)}},
\end{equation}
where $\sigma$ is a permutation of $\{1,\dotsc,n\}$ (there are obviously redundancies in this parametrization). 
Such terms can be interpreted as $n$ lines meeting, and possibly crossing, at a $2n$-valent ribbon vertex. The incoming line in position $k$ (corresponding to $A_{\alpha_k}$) crosses the vertex and go out in position $\sigma(k)$ (corresponding to $A^\dagger_{\alpha_{\sigma(k)}}$). This is pictured in figure \ref{fig:InterpretationAsLoops}. We call the drawing associated to $\sigma$ the \emph{link pattern labeled by $\sigma$}.

In this model, the most general action reads
\begin{equation} \label{matrixaction}
	S(\{A_i,A_i^\dagger\}) = \sum_{i=1}^\tau \tr A_i A_i^\dagger + \sum_{(n,\sigma)} V_{n,\sigma}(\{A_i,A_i^\dagger\}),
\end{equation}
where the sums typically run over a finite set of terms only. In the Feynman expansion, propagators connect ribbon vertices so as to form random (orientable) surfaces, as usual in matrix models. Moreover, each half-line of a ribbon vertex carries a (incoming or outgoing) line with index $i=1,\dotsc,\tau$, and these half-lines are connected by propagators to create loops. Each propagator between two vertices identifies their label $i=1,\dotsc,\tau$. As a result, there is a free sum per loop, giving rise to a factor $\tau$, hence a factor $\tau^L$ for the whole ribbon graph, $L$ being its number of loops.

The free energy of the model admits the following expansion,
\begin{equation} \label{matrixF}
	F = N^2f = -\ln \int \prod_{i=1}^{\tau}\diff A_i\diff A_i^\dagger \exp\left(-\frac{N}{\lambda}S(\{A_i,A_i^\dagger\})\right) = \sum_{\substack{\text{connected}\\\text{ribbon graphs G}}} \frac{1}{s(G)}N^{2-2g(G)}\lambda^{E-V}\tau^{L},
\end{equation}
where $s(G)$ is a symmetry factor, $E$ is the number of edges, $V$ of vertices, $F$ of faces and $L$ of loops. The $1/N$ expansion of the free energy is, as usual, the genus expansion, where the genus $g$ is
\begin{equation}
\label{eq:Genus}
	2 - 2g(G) = F - E + V.
\end{equation}

It is worth noting that two kinds of configurations may happen:
\begin{itemize}
	\item CPL configurations, where all loops are self and mutually avoiding. The name `CPL' comes from the Completely Packed Loop model. In what follows, we will see that these CPL configurations have a dominant role. They are generated by gluings of link patterns with no crossing, like on figure \ref{subfig:NonCrossingLoop}, i.e. \emph{planar} patterns up to rotations and reflection, 
	\item Configurations with crossings, where at least one loop crosses itself or another loop. 
\end{itemize}

\subsection{Mapping to colored graphs and the degree expansion of tensor models}

We will map the Feynman graphs of our matrix model to a family of edge-colored graphs which we now introduce.

\subsubsection{Colored graphs and their degree}

\begin{definition}
A $\Delta$-colored graph is a regular bipartite graph (say, with black and white vertices) where each edge carries a color from the set $\{1,\dotsc,\Delta\}$ and such that the vertices have degree $\Delta$ and the edges incidnet to a vertex all have distinct colors.
\end{definition}

\begin{figure}
\includegraphics[scale=.5]{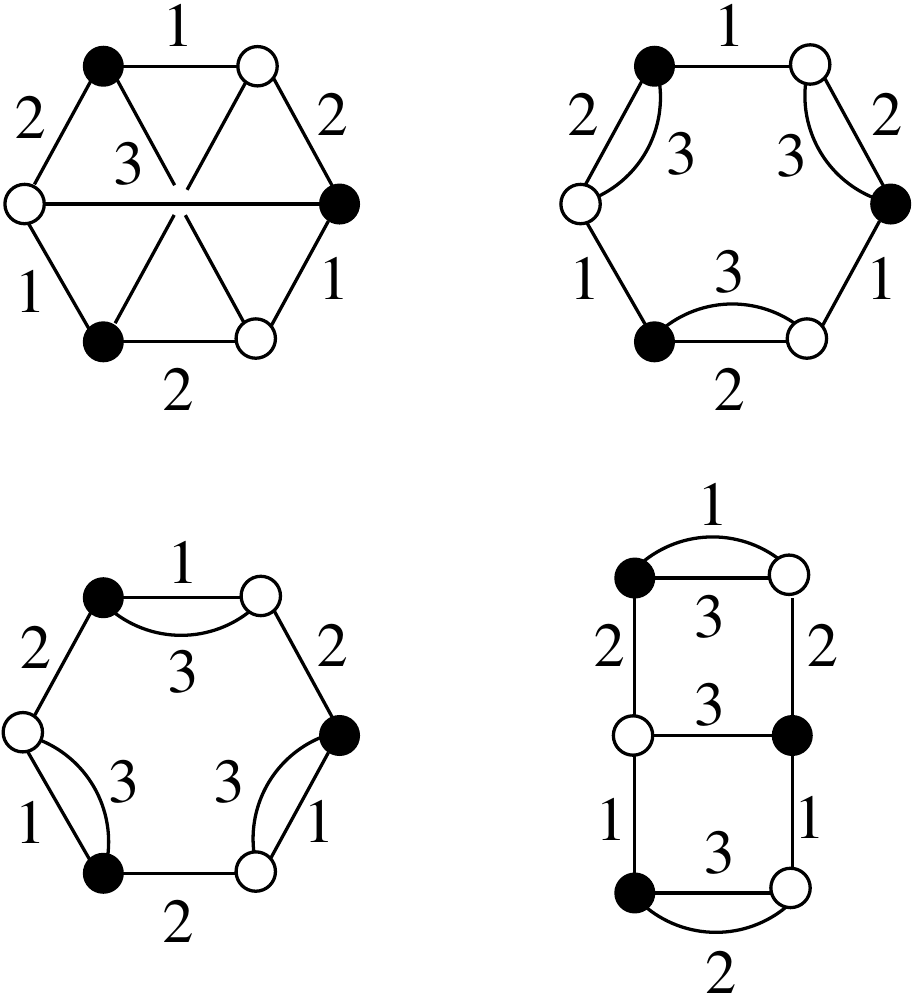}
\caption {\label{fig:bubbles} Graphs on six vertices with 3 colors.}
\end{figure}

Some graphs are given in figure \ref{fig:bubbles}. If $2p$ denotes the number of vertices in such a graph, the total number of edges is $\Delta p$, and the number of edges of any given color is simply $p$. Furthermore, coloring gives an additional structure, which provides in particular a natural notion of faces. A \emph{face with colors} $(a,b) \in \{1,\dotsc,\Delta\}$ is a closed path with alternating colors $a$ and $b$. The total number of faces of a graph $G$ is $F(G) = \sum_{a<b} F_{ab}$, where $F_{ab}$ is the number of faces with colors $a,b$.

\begin{definition}
 Let $\Delta\geq 3$ be an integer, $G$ be a connected, $\Delta$-colored graph with $2p$ vertices and $\sigma$ be a cycle on $\{1,\dotsc,\Delta\}$. The jacket $J$ associated to $\sigma$ is the connected ribbon graph which contains all the faces of colors $(\sigma^q(1),\sigma^{q+1}(1))$ for $q=0,\dots,\Delta-1$ in $G$. Therefore the number of faces in $J$ is given by $f_J = 2-2g_J + \Delta p -2p$, where $g_J$ is the genus of $J$. We define the degree $\omega(G)\in\NN$ of $G$ as the sum of the genera of the jackets.
\end{definition}

One gets an (over-)counting of faces by summing the formulas of the genus over all jackets, leading to the following theorem.

\begin{theorem} \label{thm:degree}
Let $G$ be a $\Delta$-colored graph with $2p$ vertices. The number of faces and vertices are related to the degree as follows,
\be \label{degree formula}
F - \frac{(\Delta-1)(\Delta-2)}{2}\,p = \Delta-1 -\frac2{(\Delta-2)!}\,\omega(G).
\ee
\end{theorem}

For a 3-colored graph, $2-2\omega(G) = F-p = F - 3p +2p$, where $3p$ is the total number of lines. Therefore the degree then reduces to the well-known formula of the genus. The degree was introduced for 4-colored graphs in \cite{Gur3}, and generalized in \cite{Gur4}.

\begin{figure}
\includegraphics[scale=.5]{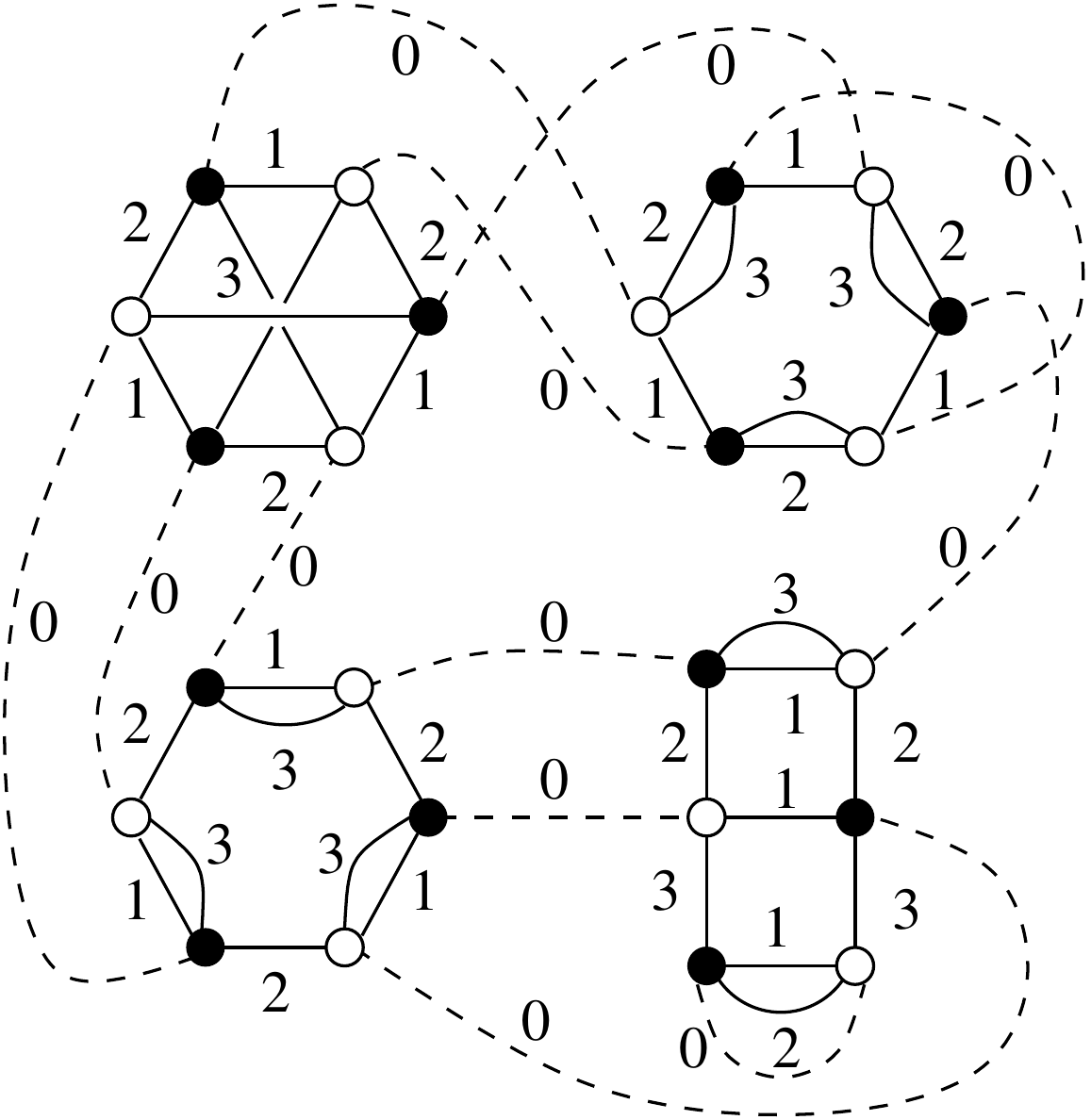}
\caption{\label{fig:feynmangraph} This is a $(3+1)$-colored graphs, obtained by connecting bubbles (in solid lines) via propagators (dashed lines, with the color 0).}
\end{figure}

The colored graphs generated by a model of a single random tensor of rank $d$, $T_{a_1 \dotsb a_d}$, are obtained from the following Feynman rules. A \emph{bubble} is a connected $d$-colored graph, with colors $1,\dotsc,d$, like in figure \ref{fig:bubbles}. It generalizes the notion of ribbon vertex used in two dimensions \cite{uncoloring}. Propagators then create edges which connect black vertices to white vertices. We assign the color $0$ to these edges. Each vertex thus receives an edge of color 0 in addition to the $d$ other edges of its bubble. Therefore the Feynman graphs are $(d+1)$-colored graphs with colors $\{0,\dotsc,d\}$ built by gluing some bubbles, as in figure \ref{fig:feynmangraph}.

Bubbles are colored graphs and therefore have a degree. Applying the degree formula \eqref{degree formula} to a Feynman graph $G$ with $d+1$ colors and to all its bubbles $\{B_\rho\}$, it comes
\be \label{graph degree}
\sum_{a=1}^d F_{0a} - (d-1)(p-b) = d - 2\left[ \frac{1}{(d-1)!}\,\omega(G) - \frac{1}{(d-2)!}\sum_\rho \omega(B_\rho)\right].
\ee
The quantity into square brackets is a positive integer, which is zero if and only if $\omega(B_\rho)=0$ for each bubble together with $\omega(G)=0$.

\begin{figure}
\includegraphics[scale=0.5]{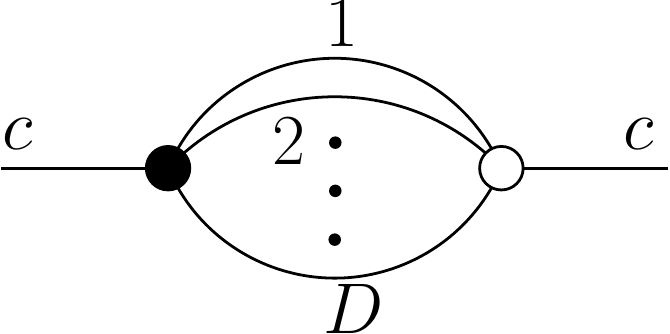}
\hspace{3cm}
\includegraphics[scale=.35]{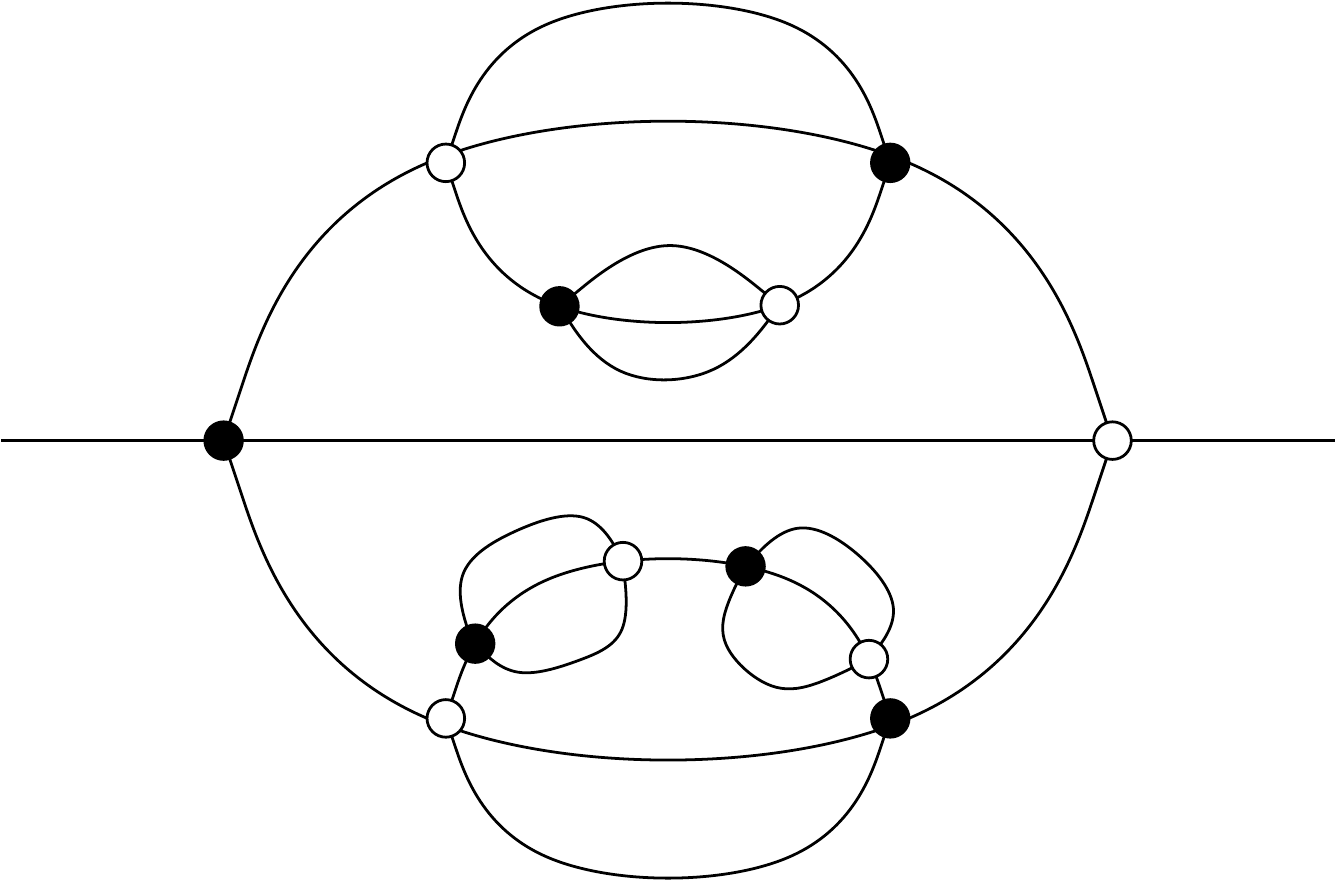}
\caption{\label{fig:dipole} On the left is a $(D-1)$-dipole with external color $c$. A 2-point (i.e. with two open half-edges) melonic graph on 4 colors is represented on the right. It is built by recursive insertions of 3-dipoles. A closed graph is obtained by connecting the two external open half-edges.}
\end{figure}

The large $N$ limit of tensor models is dominated by graphs which maximize the number of faces at fixed number of vertices and bubbles. They are therefore the graphs whose degrees vanish as well as the degrees of their bubbles. Such bubbles and Feynman graphs are called \emph{melonic}.
\begin{definition}
	A closed melonic graph with $\Delta$ colors is built by recursive insertions of $(\Delta-1)$-dipoles, i.e. two vertices connected by $\Delta-1$ lines inserted on any line, starting from the closed graph on two vertices. The $(\Delta-1)$-dipole is represented in figure \ref{fig:dipole}, as well as a melonic graph.
\end{definition}

\begin{theorem}
The colored graphs of degree $\omega=0$ are the melonic graphs.
\end{theorem}
This theorem was proved in \cite{critical-colored}.

We say that a melonic graph has melons only on the colors $a_1,\dotsc, a_k$ if it can be constructed by dipole insertions on edges of colors $a_1,\dotsc,a_k$ only.

Edge-colored graphs have been recently classified according to their degree in \cite{GurauSchaeffer}.

\subsubsection{The corresponding tensor model}

In this section, we explain why our matrix model can be written as a tensor model. Note that not all multi-matrix models are tensor models in disguise, since the action of a tensor model for a tensor of size $N_1\times \dotsm \times N_d$ is required to be $U(N_1)\times \dotsm U(N_d)$-invariant, as we explain.


Let $T_{a_1 \dotsb a_d}$ be the entries of a tensor $T$, with $a_i = 1,\dotsc,N_i$ for $i=1,\dotsc,d$, and $\overline{T}_{a_1\dotsb a_d}$ for its complex conjugate. The algebra of polynomials in the entries of $T$ and $\overline{T}$ which are invariant under the fundamental action of $U(N_1)\times \dotsm \times U(N_d)$, that is
\begin{equation} \label{U(N)Transfo}
T_{a_1\dotsb a_d} \mapsto \sum_{b_1, \dotsc, b_d} U^{(1)}_{a_1 b_1}\,\dotsm\, U^{(d)}_{a_d b_d}\ T_{b_1\dotsb b_d},
\end{equation}
where $U^{(1)},\dotsc,U^{(d)}$ are $d$ independent unitary matrices (of different sizes), and similarly for the complex conjugate $\overline{T}$, is generated by a set of polynomials labeled by bubbles. Recall that bubbles here are connected, edge-colored graphs with $d$ colors. The correspondence between invariant polynomials and bubbles works as follows. Let $B$ be a bubble. To each white (respectively black) vertex of $B$ we associate a $T$ (respectively $\overline{T}$). An edge of color $c\in\{1,\dotsc,d\}$ between a white vertex and a black vertex means that the indices in the position $c$ of the corresponding $T$ and $\overline{T}$ are identified and summed over (from $1$ to $N_c$). The polynomial labeled by $B$ can be written explicitly. Let $\mathcal{W}$ be the set of white vertices, $\mathcal{B}$ the set of black vertices and $\mathcal{E}$ the set of edges. We identify an edge $e\in\mathcal{E}$ via the black vertex and the white vertex it connects and its color, thus $e = (w,b,c)$ with $w\in\mathcal{W}, b\in\mathcal{B}$. Then the polynomial, denoted $B(T,\overline{T})$, is
\be
B(T, \overline{T}) = \prod_{k\in\mathcal{W}} \sum_{i_1^k, \dotsc, i_d^k}\ \prod_{l\in\mathcal{B}} \sum_{j_1^l, \dotsc, j_d^l}T_{i_1^k \dotsb i_d^k}\ \overline{T}_{j_1^l \dotsb j_d^l}\ \prod_{e=(w,b,c)\in\mathcal{E}}{\delta_{i_c^w,j_c^b}}.
\ee

In order to write the matrix action \eqref{matrixaction}, defined from the set of matrices $\{A_i,A^\dagger_i\}_{i=1,\dotsc,\tau}$, as a tensor action, we make the quite obvious ansatz,
\be
T_{a_1 a_2 a_3} = \left(A_{a_3}\right)_{a_1 a_2},\qquad \overline{T}_{a_1 a_2 a_3} = \left(A^\dagger_{a_3}\right)_{a_2 a_1}.
\ee
It remains to check that the matrix potential has the required invariance, that is that each $V_{n,\sigma}(\{A_i,A_i^\dagger\})$ defined in \eqref{eq:TraceInvariant}, is actually an invariant polynomial labeled by a 3-colored bubble. The above definitions of $T$ and $\overline{T}$ shows that the matrix element of a product $(A_a A_b^\dagger)_{a_1 b_1}$ is exactly $\sum_{a_2}T_{a_1 a_2 a} \overline{T}_{b_1 a_2 b}$, i.e. a contraction along the second index. Similarly, $(A_b^\dagger A_a)_{b_2 a_2}=\sum_{a_1} \overline{T}_{a_1 b_2 b} T_{a_1 a_2 a}$ is a contraction on the first index. Finally, $\sum_{i=1}^\tau (A_i)_{a_1 a_2} (A_i^\dagger)_{b_2 b_1} = \sum_{a_3} T_{a_1 a_2 a_3} \overline{T}_{b_1 b_2 a_3}$ creates the contraction along the third index. 
The action \eqref{matrixaction} is thus really a sum over invariant polynomials labeled by bubbles, and the quadratic part is obviously $\sum_{i=1}^\tau \tr A_iA_i^\dagger = \sum_{a_1,a_2,a_3} T_{a_1 a_2 a_3} \overline{T}_{a_1 a_2 a_3}$.

\begin{figure}
	\center
    \includegraphics[height=4cm]{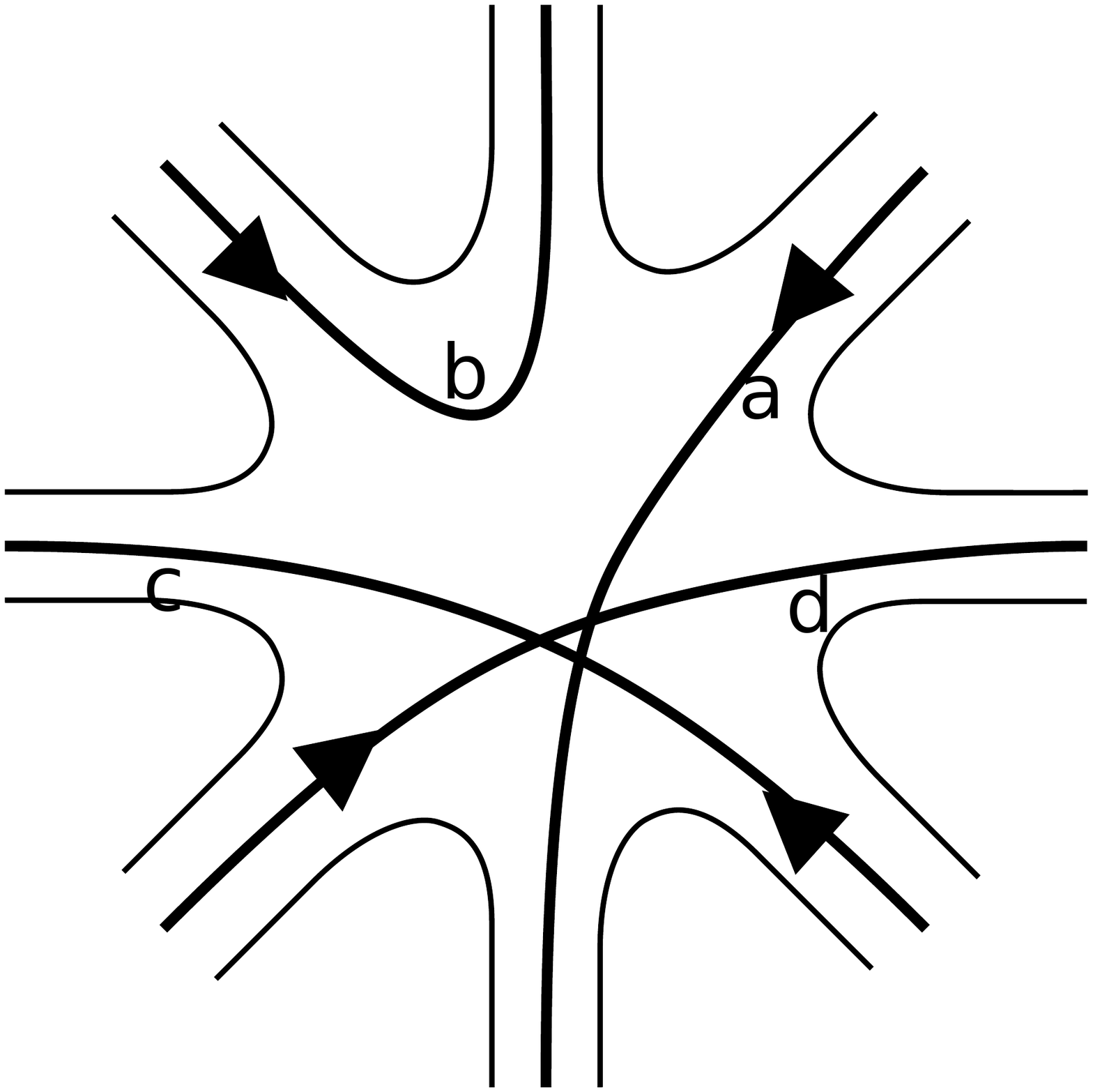}
    \hspace{0.5cm}
    \raisebox{2cm}{$\to$}
    \hspace{0.5cm}
    \includegraphics[height=4cm]{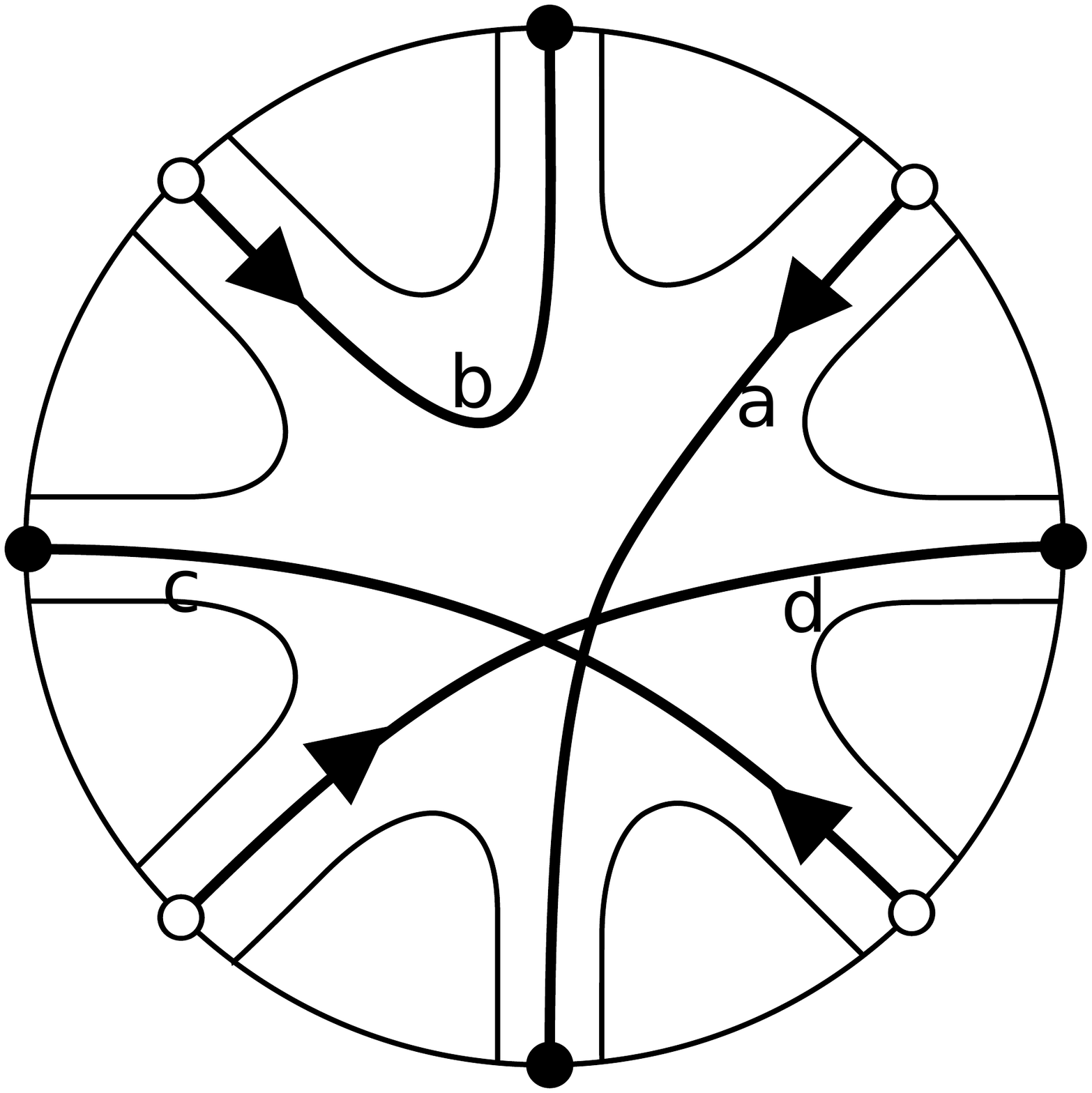}
    \hspace{0.5cm}
    \raisebox{2cm}{$\to$}
    \hspace{0.5cm}
    \includegraphics[height=4cm]{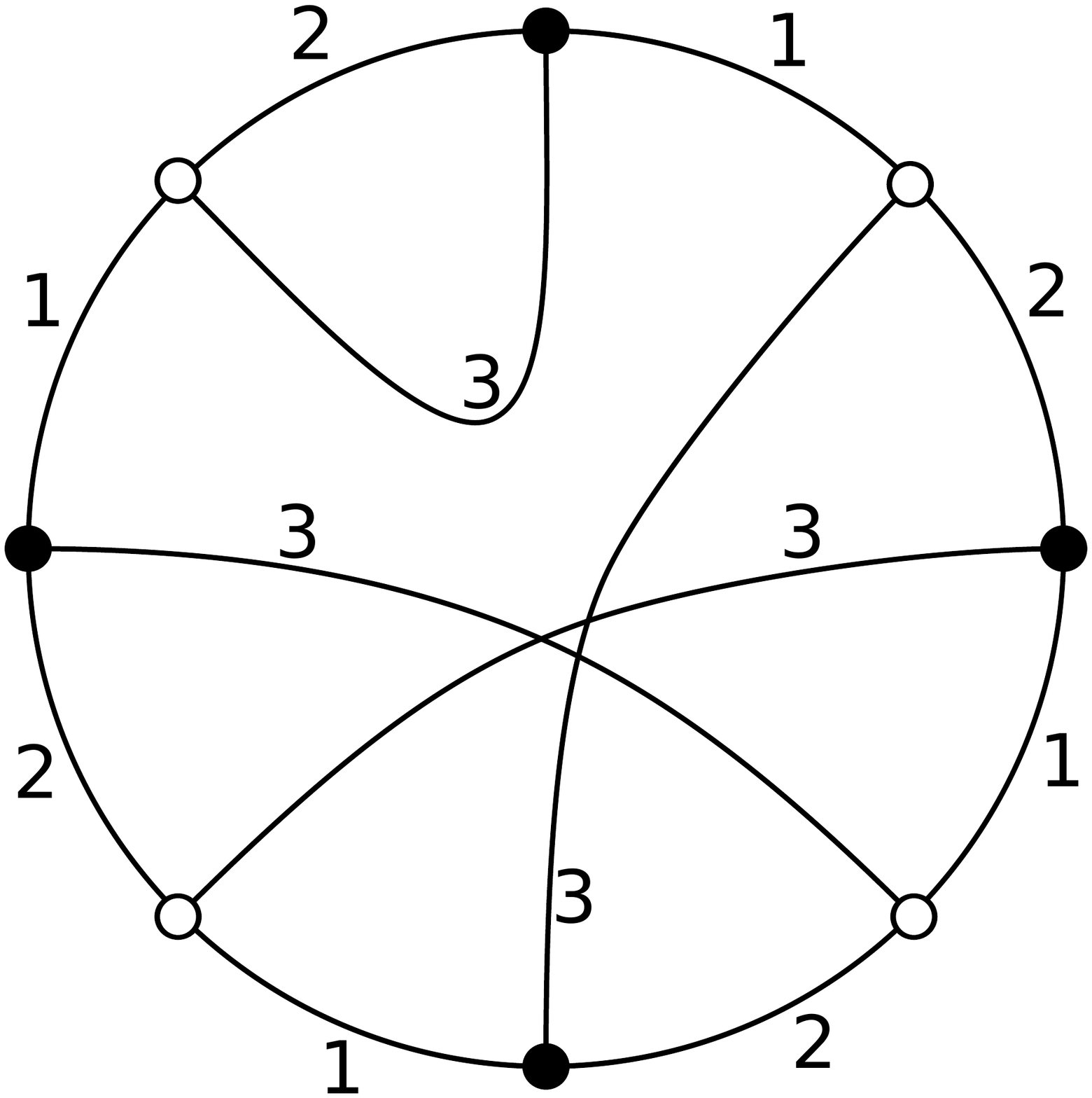}
    \caption{The map from ribbon vertices with loop lines to 3-colored bubbles. \label{fig:LoopToTensor}}
    \end{figure}

It is also interesting to proceed graphically. There is a straightforward mapping between the link patterns, i.e. the ribbon vertices of the matrix model, and bubbles with 3 colors and a single face of colors $(1,2)$, as shown in figure \ref{fig:LoopToTensor}. One draws an (unknotted) circle around the ribbon vertex, such that the intersections between the circle and the loop lines (labeled $1,\dotsc, \tau$) give rise to the vertices of the bubble (say an outgoing line gives a white vertex, and an incoming line gives a black vertex). The segments on the circle are given alternating colors 1 and 2. There are two possible choices to do that, and we choose the color 1 when going clockwise from a black to a white vertex. The loop lines which cross the ribbon vertex are then given the color 3, and they indeed connect white to black bubble vertices.

Notice that our ribbon vertices do not generate all 3-colored bubbles, but only those with a single face of colors $(1,2)$. This is because they come from single trace invariants in the matrix model.

Conversely, given a 3-colored bubble with a single face of colors $(1,2)$, one gets a unique ribbon vertex with loop lines. The ribbon vertex is determined by the face with colors $(1,2)$: there is one open ribbon line per vertex of the bubble. Each line of color 3 connects a black and a white bubble vertex and corresponds to an oriented loop line going through the ribbon vertex.

The planar patterns, used to build CPL configurations, are exactly the bubbles which are melonic with melonic insertions on the colors 1 and 2 only. This set of bubbles has been studied in \cite{SDEs} where it is shown to be in one-to-one correspondence with non-crossing partitions of $\{1,\dotsc,p\}$ up to rotations and reflections.

As an example, those on 4 vertices correspond to the terms $V_{n,\sigma}$ with $n=2$. There are only two permutations on two elements, hence two 3-colored graphs with a single face of colors $(1,2)$ which we denote $B_1$ and $B_2$, and corresponding to the identity $\sigma=(1)(2)$ and the transposition $\sigma=(12)$ (in cycle notation). Graphically, we have
\begin{equation} \label{QuarticBubbles}
\begin{aligned}
&V_{n=2,\sigma=(1)(2)} = \sum_{i,j=1}^\tau \tr( A_i A_i^\dagger A_j A_j^\dagger) = \begin{array}{c}\includegraphics[scale=.25]{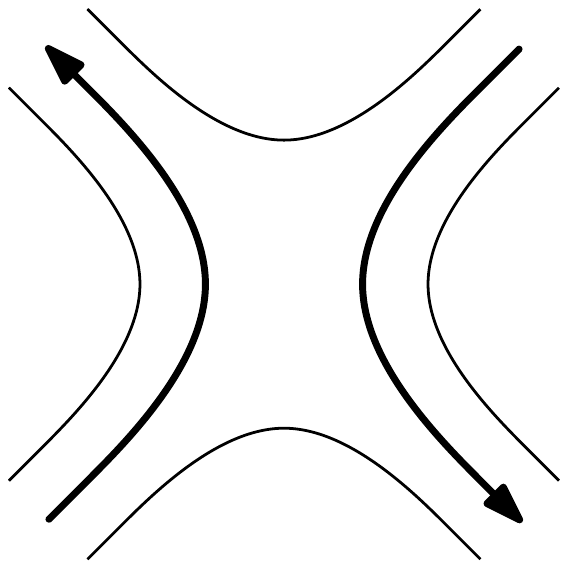}\end{array} = \begin{array}{c}\includegraphics[scale=.45]{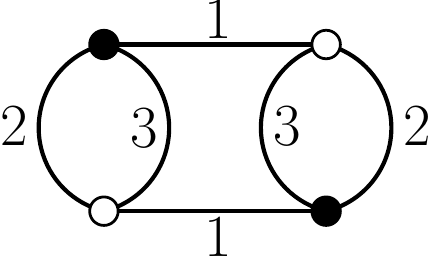}\end{array},\\
&V_{n=2,\sigma=(12)} = \sum_{i,j=1}^\tau \tr( A_i A_j^\dagger A_j A_i^\dagger) = \begin{array}{c}\includegraphics[scale=.25]{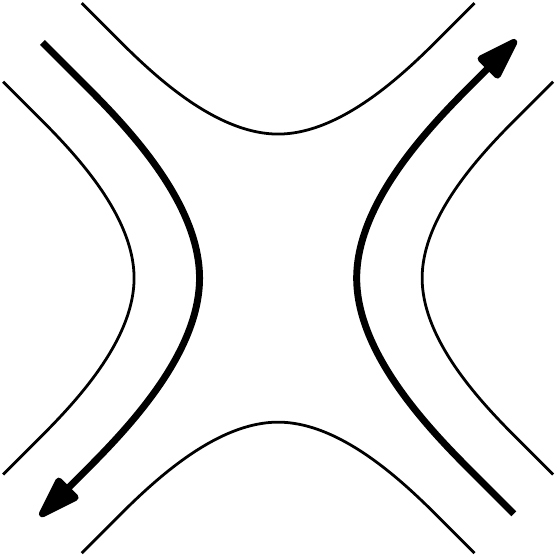}\end{array} = \begin{array}{c}\includegraphics[scale=.45]{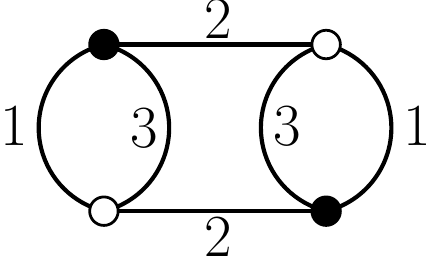}\end{array}.
\end{aligned}
\end{equation}

Therefore, we get a tensor model whose free energy expansion is \eqref{matrixF}. This implies that if we could solve exactly the matrix models defined by the potentials \eqref{eq:TraceInvariant}, we would in fact get the exact solution of some rank-three tensor models, with tensors of size $N\times N\times \tau$ (for which the $1/N$ expansion is organized according to the genus of subgraphs with colors 0,1,2 while the degree partially controls the $1/\tau$ expansion, as we are going to see).

\subsubsection{The mapping}

The fact that the initial matrix model fits in the frame of tensor models suggests the existence of a bijection between the random surfaces decorated with loops and the $(3+1)$-colored Feynman graphs of the tensor model. From the correspondence between the ribbon vertices $V_{n,\sigma}$ and the bubbles established above, this bijection is quite trivial: only propagators have to be added. The edges of the ribbon graphs are simply mapped to edges of color 0, which indeed connect black to white vertices. 

This allows to evaluate the number of faces, edges, vertices and loops of the matrix Feynman graphs in terms of faces, vertices and bubbles of the corresponding 4-colored graphs, as summarized in the Table \ref{tab:MatrixToTensorCorrespondence}.

\begin{table}
	\begin{center}
		\begin{tabular}{|c c c|}
			\hline
				Loops model graphs  & \vline & 4-colored graphs \\
			\hline
				Faces: $F$		& $\to$ & Faces of colors (0,1), (0,2): $F_{01}+F_{02}$\\
				Edges: $E$		& $\to$ & Vertexes: $p$\\
				Vertexes: $V$	& $\to$ & Bubbles of color (1,2,3): $b$\\
				Loops: $L$		& $\to$ & Faces of colors (0,3): $F_{03}$\\
			\hline
		\end{tabular}
	\end{center}
	\caption{The correspondence between the characteristics of the random surfaces with loops and the characteristics of the corresponding colored graphs. \label{tab:MatrixToTensorCorrespondence}}
\end{table}

The slightly non-trivial is to find a relation between the genus of the random surfaces and the degree of the colored graphs. Being 3-colored, the bubbles themselves can be interpreted as ribbon graphs. This is done by clockwise-ordering the edges of colors 1,2,3 around each white vertex, and counterclockwise-ordering them around each black vertex (then thickening the edges if desired). The degree of a bubble is then the genus of that discrete surface, $\omega(B) = g(B)$, and it is a measure of the amount of crossing of the loop lines at a given ribbon vertex. Obviously, the melonic bubbles are the planar ones.

Using this correspondence, the degree of a $(3+1)$-colored graph, equation \eqref{graph degree}, reads
\begin{equation}
	F + L - 2(E - V) = 3 - \omega(G) +2\sum_\rho g(B_\rho),
\end{equation}
Since the genus of the ribbon graph fixes the number of faces at fixed number of edges and vertices, through equation \eqref{eq:Genus}, the number of loops $L$ can be extracted as a function of the degree, of the genus of the subgraph of colors (0,1,2), the genera of the bubbles, and of the numbers of edges and vertices
\begin{equation}
\label{eq:LoopCounting}
	L = E - V + 1 + 2g(G) - \omega(G) + 2\sum_\rho g(B_\rho).
\end{equation}
This \emph{loop counting formula} is the main outcome of the mapping. To complete the loop counting, we prove that the quantity $\omega(G) - 2\sum_\rho g(B_\rho) - 2g(G) \geq 0$, and identify the configurations for which it vanishes. We use the obvious bound $L<F$, which together with \eqref{eq:LoopCounting} implies that
\be
\omega(G) - 2\sum_\rho g(B_\rho) - 2g(G) > -1+2g(G).
\ee
This means that if $g(G)\geq 1$, then the left hand side is strictly positive. Only the case $g(G)=0$ remains. Then we find
\be
\omega(G) - 2\sum_\rho g(B_\rho) = \frac13\,\omega(G) +\frac23 \bigl(\omega(G) - 3\sum_\rho g(B_\rho)\bigr).
\ee
In addition to $\omega(G)\geq 0$ as part of the Theorem \ref{thm:degree}, it can be proved that $\omega \geq 3\sum_\rho g(B_\rho)$, which is a particular case of the Lemma 7 in \cite{tree-algebra}.

We summarize the consequences of this analysis in the following proposition.
\begin{proposition} \label{prop:LoopCounting}
The number of loops on a random discrete surface $G$ decorated with oriented loops visiting all edges once (and such that the orientations on the edges around each vertex alternate), made of the gluing of $V$ link patterns $\{B_\rho\}_\rho$ via $E$ edges satisfies
\begin{equation*}
	L = 1 + E - V + 2g(G) - \omega(G) + 2\sum_\rho g(B_\rho),
\end{equation*}
where $\omega(G)$ is the degree of the corresponding $(3+1)$-colored graph. Furthermore
\begin{itemize}
\item The graphs of degree zero are those which maximize the number of loops at fixed number of edges and vertices (and the link pattern at each vertex is planar).
\item They are planar, $\omega(G)=0\ \Rightarrow\ g(G)=0$.
\item At fixed genus, fixed numbers of edges and vertices of each allowed type, the degree measures how far $G$ is from the configuration which maximizes the number of loops.
\end{itemize}
\end{proposition}


We also note that the formula \eqref{eq:LoopCounting} can be used to get a bound on the maximal degree of the colored graphs built from 3-colored bubbles with a single face of colors 1,2. Since $L\geq1$, it comes (using the notation of colored graphs)
\be
\omega(G)\leq p-b +2g(G) + 2\sum_\rho g(B_\rho).
\ee

Let us illustrate a bit the melonic sector in the special case where only the terms with $n=2$ are kept in the action \eqref{matrixaction}. This leaves only the two link patterns, or the two 3-colored graphs (both planar), of the equation \eqref{QuarticBubbles}. In this model, all melonic insertions come from inserting appropriately one of these two bubbles on any edge of color 0. Assuming the edge of color 0 correponds to a ribbon edge with a loop going up north, the two possibilities are
\begin{equation}
\begin{array}{c}\includegraphics[scale=.5]{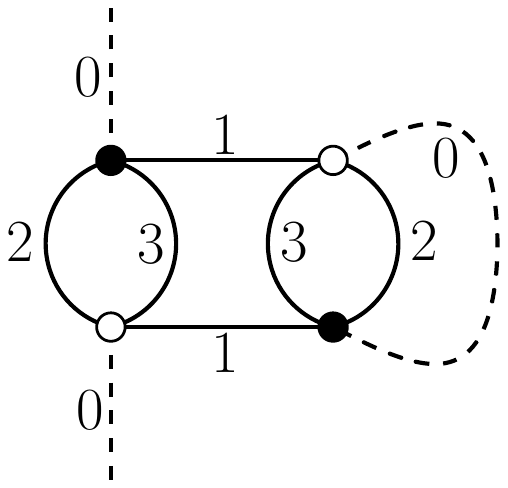} \end{array} = \begin{array}{c}\includegraphics[scale=.25]{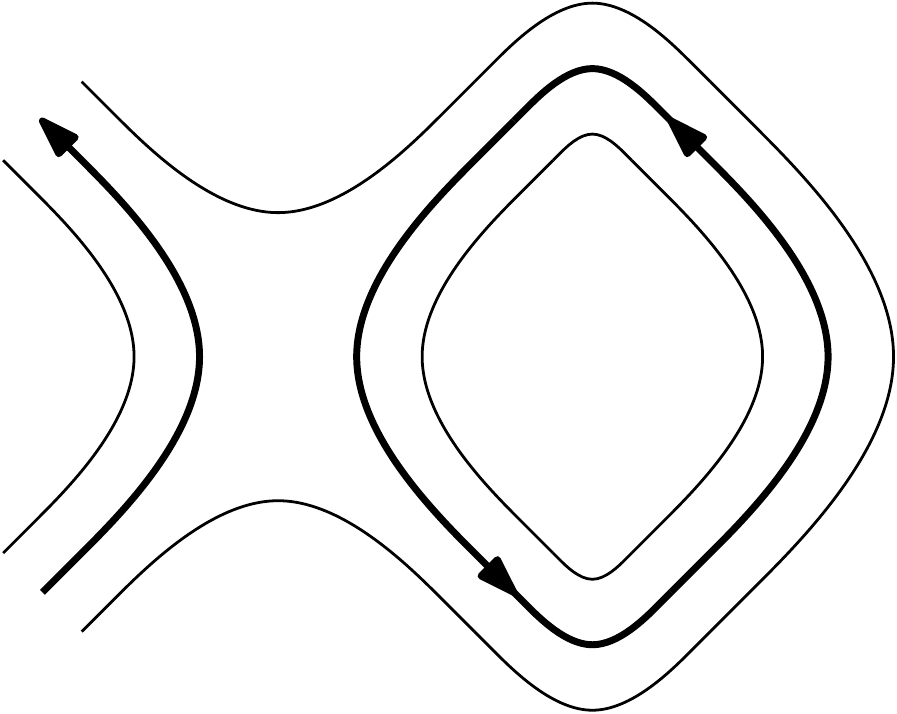} \end{array}\qquad \text{and}\qquad 
\begin{array}{c}\includegraphics[scale=.5]{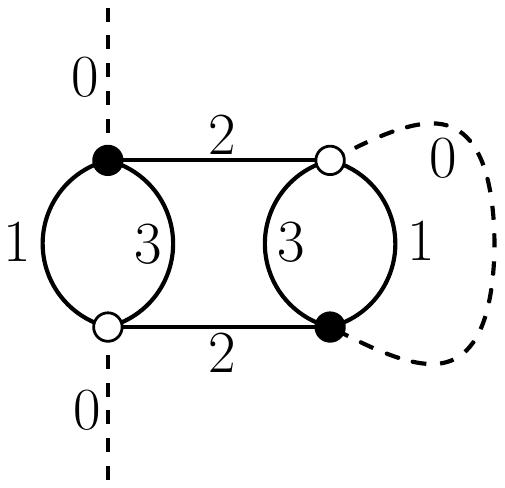} \end{array} = \begin{array}{c}\includegraphics[scale=.25]{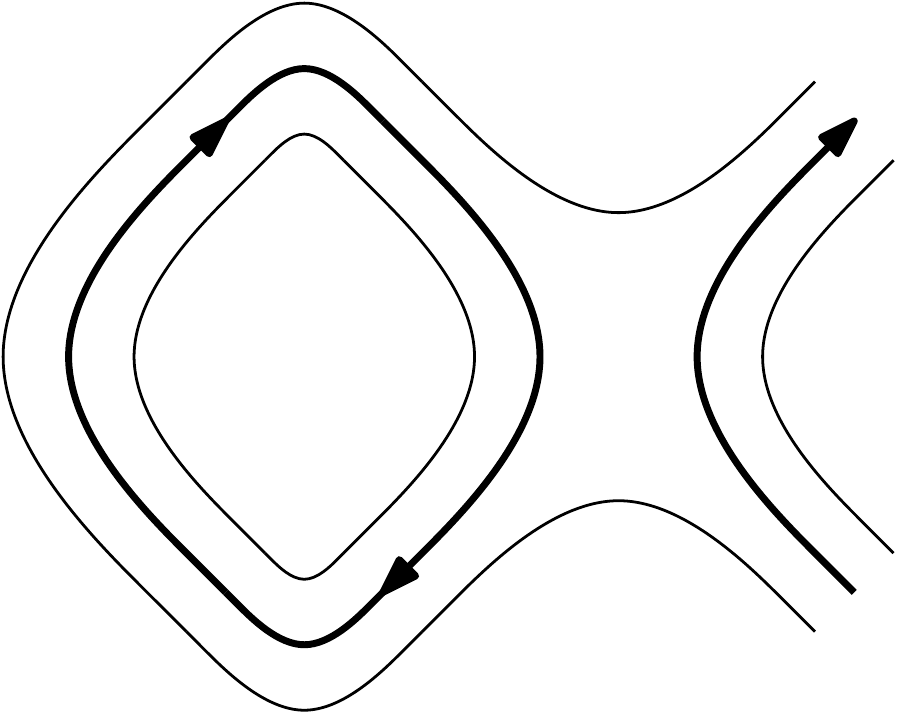} \end{array}.
\end{equation}
We see that a melonic insertion adds one loop, one ribbon vertex and one ribbon edge (and does not change the genus). By contrast, a non-melonic insertion on an edge of color 0 would be
\begin{equation}
\begin{array}{c}\includegraphics[scale=.5]{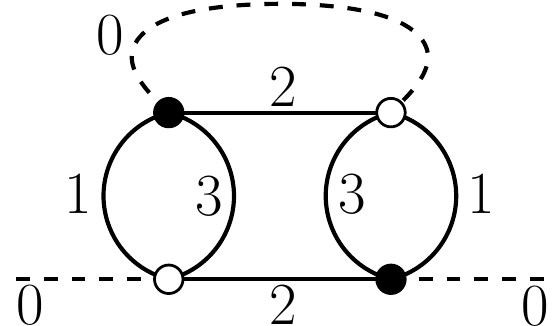} \end{array} = \begin{array}{c}\includegraphics[scale=.25]{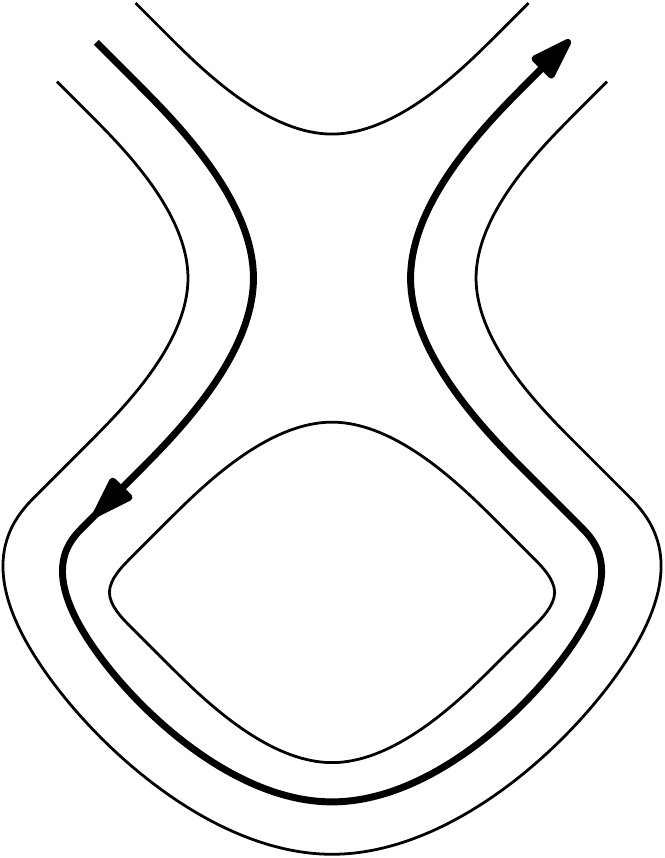} \end{array}
\end{equation}
which would not create a new loop.

In the section which follows, we solve explicitly the melonic sector ($\omega=0$) and further use the recent classification of colored graphs \cite{GurauSchaeffer} to organize the $1/\tau$ expansion.

\subsection{Scaling limits} \label{sec:ScalingLimits}

Using the counting of loops obtained in equation \eqref{eq:LoopCounting}, the free energy writes
\begin{equation}
	N^2 f = \sum_{\substack{\text{connected}\\ \text{ribbon graphs}}} \left(\frac{N}{\tau}\right)^{2-2g(G)}\ \left(\lambda \tau\right)^{E-V}\ \tau^{3-\omega(G)}\ \frac1{s(G)}.
\end{equation}
This shows three contributions to the exponent of $\tau$. Those with the genus and with $E-V$ are not relevant since these quantities are controled by $N$ and $\lambda$. Consequently, the degree $\omega$ labels the expansion in the number of loops.

\subsubsection{Large $\tau$ limit}

Furthermore, it is possible to build a scaling limit which projects the loop model onto the melonic sector. To project onto the melonic family, the limit $\tau \to \infty$ is required. To ensure the limit is well-defined, we must scale $\lambda$ with $\tau$ as follows: $\lambda\tau = \tilde{\lambda}$, where $\tilde{\lambda}$ is kept finite. We also scale $N$ with $\tau$, and for convenience set their ratio to 1, $\tau = N$. The rescaled free energy $\tilde{f} = \frac{f}{\tau} =  \frac{f}{N}$ then reads
\begin{equation} \label{FreeEnergyTensor}
	N^3 \tilde{f} = N^2 f = \sum_{\substack{\text{4-colored}\\\text{connected graphs}}} N^{3-\omega(G)} \tilde{\lambda}^{E-V} \frac1{s(G)},
\end{equation}
It is finite in the large $N$, large $\tau$ limit, and its leading order in the $1/N$ expansion consists of melonic graphs.

It is interesting to perform the rescaling directly in the matrix integral (and setting $\tau$ to $N$ everywhere),
\begin{equation} \label{tensorscaling}
	N^3 \tilde{f} = -\ln \int \prod_{i=1}^N \diff A_i \diff A_i^\dagger \exp\left( -\frac{N^2}{\tilde{\lambda}} S(\{A_i,A_i^\dagger\})\right)
\end{equation}
The factor $N^2$ in front of the action is exactly the standard scaling for a random tensor of rank-three and size $N^3$. This is natural in this scaling limit, since there are $\tau=N$ matrices, each of size $N\times N$.

We can write the solution quite explicitly in the large $N$ limit \cite{universality, uncoloring}. Indeed, large random tensors in a unitary-invariant distribution (invariant under \eqref{U(N)Transfo}) are subjected to a universality theorem, stating that all large $N$ expectations are Gaussian, with the covariance being the large $N$ 2-point function. For an invariant polynomial $B(T,\overline{T})$ of degree $p_B$ in $T$, this gives
\begin{equation}
\frac1N\,\langle B(T,\overline{T})\rangle = N^{-\omega^*(B)} \Bigl(C_B\ G_2^{p_B}+\mathcal{O}(1/N)\Bigr),
\end{equation}
where $G_2 = \lim_{N\to\infty} \langle T\cdot\overline{T}\rangle/N = \lim_{N\to\infty} \langle \sum_{i=1}^N \tr (A_iA_i^\dagger)\rangle/N$. $\omega^*(B) \geq 0$ and vanishes if and only if $B$ is melonic, meaning that melonic bubbles of the action are the only relevant ones at large $N$. Therefore, only the terms of the type $V_{n,\sigma}$ in \eqref{matrixaction} with $\sigma$ corresponding to a planar link pattern survive. Moreover, $C_B$ is the leading order number of Wick contractions and for a melonic bubble turns out to be 1 only. This way,
\begin{equation}
\frac1N\,\langle V_{n,\sigma \rm{ planar}}(\{A_i,A_i^\dagger\}) \rangle = G_2^n.
\end{equation}
All large $N$ calculations thus boil down to the leading order 2-point function. It is found thanks to the Schwinger-Dyson equation
\begin{equation}
\sum_{a_1,a_2,i=1,\dotsc,N} \int \prod_i dA_i\,dA_i^\dagger\ \frac{\partial}{\partial (A_i)_{a_1 a_2}} \Bigl((A_i)_{a_1 a_2}\ e^{-N^2 S(\{A_i,A_i^\dagger\})/\tilde{\lambda}}\Bigr) = 0,
\end{equation}
which after making the derivatives explicit and using the universality to close the system leads to the equation
\begin{equation}
\tilde{\lambda} - G_2 - \sum_{n, \text{planar }\sigma} n\,G_2^n = 0,
\end{equation}
which is polynomial as long as the action contains a finite collection of planar link patterns. It is a standard result that one then gets a square-root singularity for $G_2$ when approaching the critical value of $\tilde{\lambda}$, i.e. $G_2 \sim (\tilde{\lambda}_c - \tilde{\lambda})^{1/2}$. Therefore the singular part of the free energy behaves as $\tilde{f} \sim (\tilde{\lambda}_c - \tilde{\lambda})^{2-\gamma}$ with $\gamma=1/2$. The regime where $\tilde{\lambda}$ is close to $\tilde{\lambda}_c$ is called the \emph{continuum limit}.

\subsubsection{The $1/\tau$ expansion}

Thanks to the $1/N$ expansion, we can work at fixed genus. We can then take advantage of the recent classification of edge-colored graphs according to their degree \cite{GurauSchaeffer} to organize the the $1/\tau$ expansion.

This classification relies on the fact that only 2-point subgraphs and 4-point subgraphs can generate infinite family of graphs of constant degree\footnote{We remind the reader that tensor model at large $N$ are dominated by Gaussian contributions, i.e. 2-point functions, while the first $1/N$ correction only involves 2-point and 4-point functions, \cite{NLO, DSSD}.}. Once replaced by ``reduced'' 2-point and 4-point functions, there exists only a finite number of graphs of given degree. Those reduced graphs are called \emph{schemes} in \cite{GurauSchaeffer}.

In the following, we restrict the potential to $n=2$, leaving only room for the bubbles $B_1, B_2$ introduced in \eqref{QuarticBubbles}. (This reduces the source of 4-point functions; otherwise we would have for instance a 6-point bubble with an arbitrary 2-point function between two of its vertices also play the role of an effective 4-point bubble, and so on\footnote{The reference \cite{GurauSchaeffer} studies the whole set of colored graphs, which is somewhat simpler than focusing on the set of graphs generated by a given but arbitrary set of bubbles, except if this set is simple enough. This is the case for graphs built from quartic interactions ($n=2$ here), and this is the choice made in \cite{DSQuartic}.}). We recall the loop counting formula specialized to this case (i.e. with $E=2V$ and planar link patterns),
\begin{equation}
L = V +(3-\omega(G))-(2-2g(G)).
\end{equation}
Therefore the degree measures how far a loop configuration is from the one which maximizes the number of loops for a fixed number of vertices and a fixed genus.

The Schwinger-Dyson equation on $G_2$ simply reads $\lambda - G_2 - 4 G_2^2=0$, hence
\begin{equation}
G_2 = \frac{\sqrt{1+16\lambda}-1}{8},
\end{equation}
with $G_2\sim \lambda$ for small $\lambda$. The critical point which defines the continuum limit is $\lambda_c=-1/16$.

Given an arbitrary $(3+1)$-colored graph built from the bubbles of type $B_1, B_2$ glued along edges of color 0, one first reduces the purely melonic 2-point subgraphs, as the one in figure \ref{fig:dipole} (this is done recursively by identifying 2-cut edges; the order does not matter, as proved in \cite{GurauSchaeffer}). Arbitrary melonic insertions do not change the degree, and so does this reduction. This way, we have to consider only \emph{melon-free} graphs, while all melonic insertions are completely accounted for by simply using $G_2(\lambda)$ as the new propagator, i.e. $G_2$ becomes the weight associated to edges of color 0.

Second, one identifies \emph{chains}. In our model, chains are simply sequences of quartic bubbles glued in a chain-like manner,
\begin{equation*}
\begin{array}{c}\includegraphics[scale=.45]{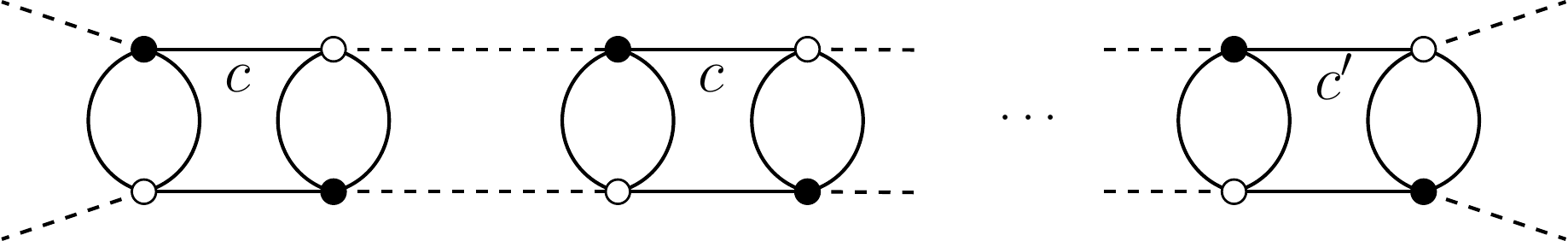}\end{array}
\end{equation*}
where $c,c'$ are 1 and/or 2. Those chains have to be maximal, so they have 4 half-edges of color 0 as external edges. There are two types of chains.
\begin{itemize}
\item Those built from a sequence of a single bubble, either $B_1$ or $B_2$, and called \emph{unbroken chains}. In terms of ribbon graphs and loops, an unbroken chain takes the form
\begin{equation*}
\begin{array}{c}\includegraphics[scale=.4]{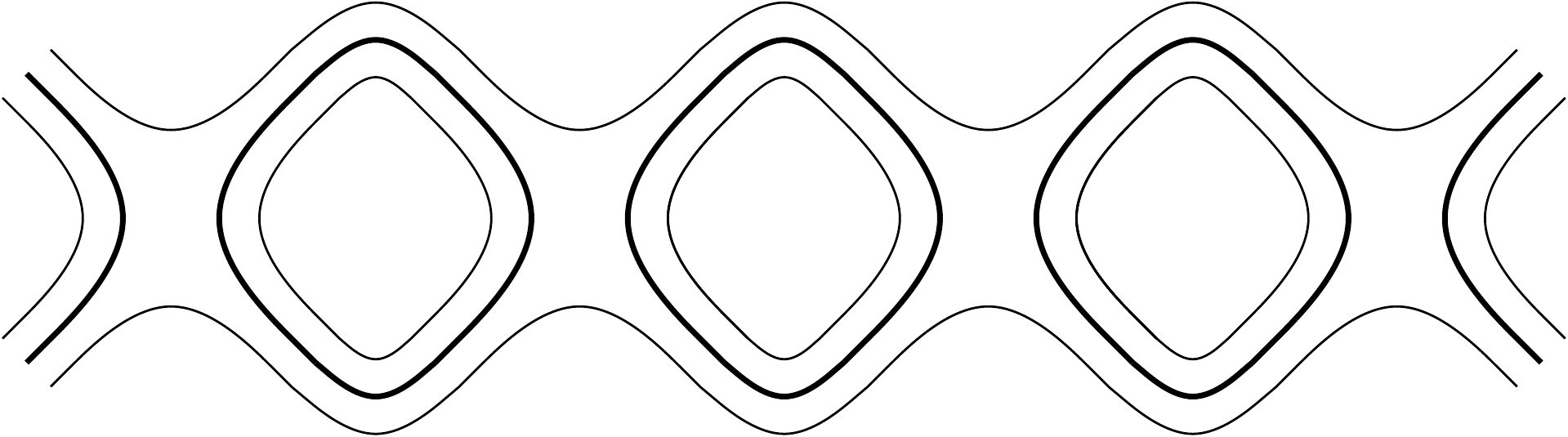}\end{array}
\end{equation*}
with two possible orientations. It is clearly planar.

The generating function of unbroken chains with a weight $-2/\lambda$ on each bubble and $G_2(\lambda)$ on each edge of color 0 is
\begin{equation}
C_u(\lambda) = \sum_{n\geq 1} \frac{(-2)^n\,(G_2(\lambda))^{2(n-1)}}{\lambda^n} = -\frac{2}{\lambda + 2(G_2(\lambda))^2}.
\end{equation}

\item Those which contain both bubbles $B_1, B_2$ and called \emph{broken chains}. The generating function of broken chains is found by considering the one of arbitrary chains, with arbitrary 4-point bubbles (bubbles hence receiving the weight $-2\times 2/\lambda$, to account for the two possible types of bubbles at each time), and substracting the generating functions of the two unbroken chains,
\begin{equation}
\begin{aligned}
C_b(\lambda) &= \sum_{n\geq 1} \frac{(-4)^n\,(G_2(\lambda))^{2(n-1)}}{\lambda^n} -2C_u(\lambda) = -\frac{4}{\lambda + 4(G_2(\lambda))^2} + \frac{4}{\lambda + 2(G_2(\lambda))^2} \\
&= \frac{8(G_2(\lambda))^2}{\lambda + 2(G_2(\lambda))^2}\ \frac1{\lambda + 4  (G_2(\lambda))^2}.
\end{aligned}
\end{equation}
\end{itemize}

Chains can be arbitrarily long with no change in the degree of the melon-free graphs. We have to make sure that does not change the genus of the random surface neither. It is clear for the unbroken chains. As for the broken ones, if a bubble $B_i$ is inserted somewhere in an unbroken subchain of type $i$, including at an end, this does not change anything. So we are left with the case where a bubble $B_1$, for instance, is inserted in a subchain of bubbles $B_2$. Because the full chain is broken, there is another bubble $B_1$ somewhere, say on the right of the chain,
\begin{equation}
\begin{array}{c}\includegraphics[scale=.5]{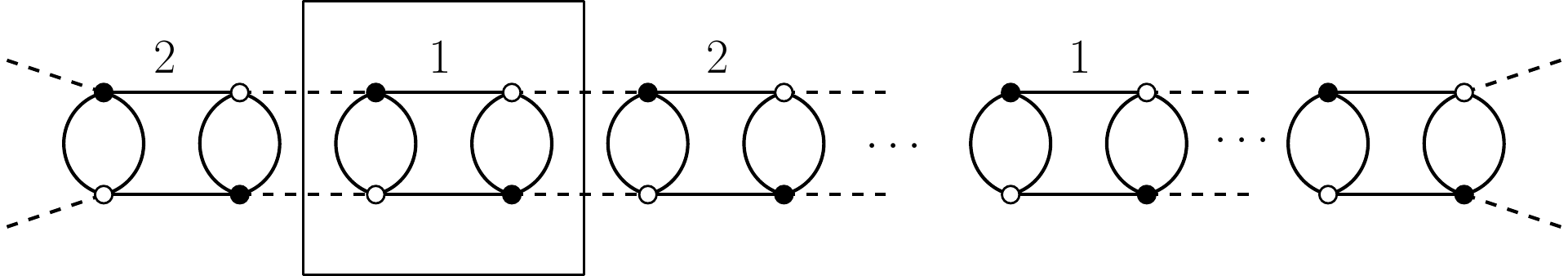}\end{array}
\end{equation}
where we have marked the added bubble in a bounding box. We have to evaluate the variation of the genus of the subgraph with colors 0,1,2 between before and after the insertion. Clearly, the number of ribbon edges changes by 2 and the number of ribbon vertices (i.e. bubbles) by 1. Therefore $\Delta (E-V) = 1$. To find the variation of the number of faces, it is more convenient to use the representation as an edge-colored graph rather than as a ribbon graph. The number of faces of the surface is $F=F_{01}+F_{02}$. The face of colors $(0,2)$ which arrives from the top left leaves on the bottom left, and that was already the case before the insertion because the chain is broken. There is however a new face of colors $(0,2)$, which goes around the bubble of type $B_2$ on the right of the bounding box. Therefore $\Delta F_{02}=1$. Moreover, there is no new face of colors $(0,1)$, so $\Delta F_{01}=0$. The variation of the genus is thus $-2\Delta g= \Delta (F-E+V) = 0$, meaning that the genus is independent of the length of the chain.

As a consequence, it is safe to simply contract chains into ``boxes'' called broken or unbroken chain-vertices with two incident edges of color 0 on one side of the box and two on the opposite side, and weight them with the generating functions $C_b(\lambda)$ or $C_u(\lambda)$ respectively. We obtain this way the set of \emph{schemes}, i.e. melon-free graphs with chain-vertices representing arbitrarily long chains. The key result of \cite{GurauSchaeffer} is then the finiteness of the number of schemes at any fixed degree.

Let $s$ be a scheme with $p\geq2$ black vertices, $\alpha$ unbroken chain-vertices and $\beta$ broken chain-vertices. Then the generating function of colored graphs, rooted on an edge of color 0, with scheme $s$ is
\begin{equation}
G_s(\lambda) = (G_2(\lambda))^{p}\,(C_u(\lambda))^\alpha\,(C_b(\lambda))^\beta.
\end{equation}
To get the free energy of the model (or rather its 2-point function), one substitutes $\lambda \tau$ instead of $\lambda$. Then the 2-point function at genus $g$ has the expansion
\begin{equation}
G_2^{(g)}(\lambda) = G_2(\lambda\tau)\,\delta_{g,0} + \sum_{\omega\geq1} \tau^{3-\omega} \sum_{\substack{\text{schemes $s$} \\ \omega(s)=\omega, g(s)=g}}  (-2)^\alpha 8^\beta \frac{(G_2(\lambda\tau))^{p+2\beta+1}}{(\lambda\tau + 2(G_2(\lambda\tau))^2)^{\alpha+\beta}\ (\lambda\tau+4(G_2(\lambda\tau))^2)^\beta},
\end{equation}
where the sum over schemes at fixed genus and degree is finite. Here we have isolated the purely melonic part, which corresponds to the empty scheme with no vertices.

Of course, to complete the analysis, it is necessary to know how the degree of a scheme behaves as a function of the number of chains. Again, this was done in \cite{GurauSchaeffer}. If a chain-vertex is separating, i.e. if after its removal and after connecting the half-edges of color 0 together on each side of the chain vertex we get two connected components, then the degree of the graph is simply the sum of the degrees of both connected components. For an non-separating, unbroken chain of type $i$, such that there are two different faces of colors $(0,i)$ going through the chain, the degree is the degree of the graph with the chain removed plus one, meaning such a chain contribute to a factor $1/\tau$. In all other situations, a chain brings in a factor $1/\tau^3$.

A corollary of this analysis is the double scaling regime. First notice that the critical point defining the continuum limit is $\lambda_\tau = -1/(16\tau)$. The generating function $C_u(\lambda\tau)$ of unbroken chains is finite at criticality. However, the generating function of broken chains is singular. Indeed, its denominator contains
\begin{equation}
\lambda\tau + 4 (G_2(\lambda\tau))^2 = -\lambda\tau\ G_2(\lambda\tau)\ \sqrt{1-\lambda/\lambda_\tau}.
\end{equation}
Therefore a scheme with $\beta$ broken chains diverges as $(1-\lambda/\lambda_\tau)^{\beta/2}$ at criticality. The idea of the double scaling limit is to pick up the terms of arbitrary degree which maximize the divergence. 

The answers provided in \cite{GurauSchaeffer} for generic edge-colored graphs and in \cite{DSQuartic} in the special case of quartic melonic interactions coincide. We consider a rooted, binary tree with a single loop attached to every leaf. For each such tree, we get an edge-colored graph of the model by replacing the edges with broken chains which are glued together at the vertices in the obvious way, while the loops on the leaves represent unbroken chains. Those loops break melonicity and were called cherries in \cite{DSQuartic}. Since all broken chains are separating, the degree is simply the number of unbroken ones, $\omega=n$. Moreover, the binary-tree structure of broken chains is the way to maximize the divergence at fixed number of cherries. The number of broken chains grows linearly as two times the number of cherries, so each such graph receives a factor $\tau^{-n}/\sqrt{1-\lambda/\lambda_\tau}^{2n}$. This shows that the optimal balance is reached by introducing $x = \tau (1-\lambda/\lambda_\tau)$ and sending $\tau\to\infty, \lambda\to\lambda_\tau$ while $x$ is kept fixed.

Using the same technique as in \cite{GurauSchaeffer} to extract the behavior of the degree as a function of the chain-vertices, the graphs of the double-scaling regime can be shown to be planar. Indeed, one breaks up the cherry trees into isolated vertices, edges which represent broken chains, and loops attached to the leaves and analyze the genus of each piece (found to vanish in all cases).

The resummation of this family is quite simple to perform\footnote{Compared to \cite{DSQuartic}, one has to set $D=3$ when $D$ enters the degree, but $D=2$ in the equations for criticality since we have only two quartic bubbles and not three.}. It has a square-root singularity in $x$ that is likely to lead to a branched polymer phase.

\subsection{Another bijection and the intermediate field method} \label{sec:Bimaps}

We have shown a bijection between the ribbon graphs with oriented loops generated by the matrix model \eqref{matrixF} and the edge-colored graphs of tensor models whose interactions are labeled by bubbles with a single cycle of colors $(1,2)$. In the case the potential is restricted to the two quartic terms in equation \eqref{QuarticBubbles}, there is a bijection between the graphs of the tensor model and a family of maps. It was first observed in \cite{BeyondPert} in quartic melonic models, and generalized to tensor models with arbitrary quartic interactions in \cite{GenericQuartic}. Algebraically, this bijection corresponds to the intermediate field method. Here, we first present the bijection, then the corresponding intermediate field theory.

Notice that the bubbles used in the quartic case, equation \eqref{QuarticBubbles}, have four vertices with a canonical partition in pairs. A canonical pair of vertices consists of those connected by a multiple edge (here two edges including the one of color 3). The two pairs are connected by two edges of color $i$ (here 1 or 2) and we have labeled the bubble by that color ($B_1$ and $B_2$). We are now going to represent $B_i$ as an edge of color $i$, as if the canonical pairs of vertices were contracted to single points.

Furthermore, each white/black vertex has an incident edge of color 0. This means that every edge of color 0 belongs to a single closed cycle made of alternating edges of color 0 and multiple edges. We map those cycles to vertices, while preserving the cyclic ordering of the bubbles. In our model, those vertices correspond to the faces of color $(0,3)$, i.e. the loops. Through this process, we represent every colored graph as a map, since the ordering around each vertex matters, with edges of colors 1 or 2.

For such a map, there are two canonical submaps, $\cM_1$ and $\cM_2$, which respectively correspond to the submaps  containing only the edges of color 1 and 2. The faces of colors $(0,i)$ in the Feynman graph of the tensor model are mapped to the faces of the map $\cM_i$. Moreover, the bubbles are mapped to edges and the loops to vertices.

Remarkably, many problems in tensor models become quite simple when formulated in this way. For instance, the dominance of the melonic sector: the question is how to maximize the number of loops at fixed number of ribbon vertices. After the mapping, it becomes how to maximize the number of vertices at fixed number of edges; the answer clearly being trees, which indeed are the representatives of the melonic edge-colored graphs. Further, the double scaling regime presented in the previous section is dominated by Motzkin trees (i.e. trees whose nodes can have zero, one or two children), such that there always is at least one change of color between two vertices of degree three, and with loops of arbitrary length and of a fixed color attached to the leaves.

Since we exhibited a bijection to maps, there may be a matrix model which generates them with the correct amplitudes. This works through the intermediate field method which transforms the initial matrix model \eqref{matrixF} with $n=2$ into a two-Hermitian-matrix model. Here it is useful to introduce independent coupling constants $\lambda_1, \lambda_2$ and consider
\begin{equation}
Z_{N,\tau}(\lambda_1,\lambda_2) = \int \prod_{i=1}^\tau dA_i\,dA_i^\dagger\ e^{-N\left(\sum_i \tr A_iA_i^\dagger + \lambda_1 \tr \sum_{i,j} A_i A_i^\dagger A_j A_j^\dagger + \lambda_2 \tr \sum_{i,j} A_i A_j^\dagger A_j A_i^\dagger\right)}.
\end{equation}
We can re-write each quartic term via a Gaussian integral over an auxiliary, Hermitian matrix,
\begin{equation}
e^{-N\lambda_1\tr \sum_{i,j} A_i A_i^\dagger A_j A_j^\dagger} = \int dM_1\ e^{-N \tr M_1^2 -2iN\sqrt{\lambda_1} \tr \sum_i M_1 A_i A_i^\dagger},
\end{equation}
up to irrelevant constants, and similarly for the other quartic term. The partition function is then
\begin{equation}
Z_{N,\tau}(\lambda_1,\lambda_2) = \int dM_1\,dM_2\,\prod_{i=1}^\tau dA_i\,dA_i^\dagger\ e^{-N\tr(M_1^2+M_2^2+\sum_i A_iA_i^\dagger) - 2iN \sqrt{\lambda_1} \tr M_1 \sum_i A_i A_i^\dagger - 2iN\sqrt{\lambda_2} \tr M_2 \sum_i A_i^\dagger A_i}.
\end{equation}
Performing the Gaussian integral on the $\tau$ matrices $A_i$, one gets,
\begin{equation} \label{TwoMatrixModel}
Z_{N,\tau}(\lambda_1,\lambda_2) = \int dM_1\,dM_2\ e^{-N\tr(M_1^2+M_2^2) - \tau \tr \ln \left(\mathbb{I}\otimes \mathbb{I} - 2i\sqrt{\lambda_1} M_1\otimes \mathbb{I} -2i\sqrt{\lambda_2}\mathbb{I}\otimes M_2\right)}.
\end{equation}
If the logarithm is expanded onto powers of $M_1, M_2$, it is clear that we have a generating function for the maps described above.

We are not aware of a solution of this model for arbitrary $\lambda_1,\lambda_2,\tau$ in the literature. Nevertheless, setting $\lambda_2=0$, one gets
\begin{equation} \label{SingleColorQuartic}
Z_{N,\tau}(\lambda_1,0) = \int \prod_{i=1}^\tau dA_i\,dA_i^\dagger\ e^{-N\left(\sum_i \tr A_iA_i^\dagger + \lambda_1 \tr \sum_{i,j} A_i A_i^\dagger A_j A_j^\dagger\right)} = \int dM\ e^{-N\tr M^2 - N\tau \tr \ln \left(\mathbb{I} - 2i\sqrt{\lambda_1} M_1\right)}.
\end{equation}
One recognizes here the generalized Penner model with a quadratic potential. We refer to \cite{PennerAllGenera} for an analysis with an arbitrary polynomial, at all genera, using the loop equations. In the case of the quadratic potential, the Penner model with coupling $\tau$ on the logarithmic part is equivalent to the quartic matrix model with $\tau$ matrices, which can actually be solved directly. For instance, a rectangular matrix of size $N\times \tau N$ can be formed, $C_{a_1 \alpha} = (A_i)_{a_1 a_2}$ with the ``fat'' index $\alpha=(a_2,i)$. Then the action is simply $\tr CC^\dagger + \lambda_1 \tr (CC^\dagger)^2$, and the partition function can be evaluated using techniques developed for rectangular matrix models, like the orthogonal polynomials in \cite{MyersTripleScaling} (and see \cite{toy-doublescaling} for an application to tensor models). 


As far as we know, the quartic case with $\lambda_2=0$ is the only situation where a model of the generic class we have introduced has been solved. However, it should be emphasized that already the quartic model with $\lambda_2\neq 0$ is very different. In particular, for $\lambda_2=0$, the distinction between broken and unbroken chains disappears and all chains become singular at the critical point. 

\section*{Conclusion}

The motivation of this article is to connect tensor models and its challenges to the more familiar framework of matrix models.

With this in mind, the present article has been devoted to a novel presentation of random tensor models, from the view of matrix models. It is based on a really simple observations: that a tensor of size $N\times N \times \tau$ can be seen as a set of $\tau$ matrices. 

In section \ref{sec:loops}, this observation allows to interpret models for tensors of size $N\times N\times \tau$ whose interactions have a single cycle of colors $(1,2)$ as $U(\tau)$-invariant matrix models. We describe this correspondence through a bijection between edge-colored graphs and random surfaces decorated with oriented loops and show that the degree, which organizes the $1/N$ expansion of tensor models, here organizes the expansion with respect to the number of loops on the random surfaces, via the equation \eqref{eq:LoopCounting}. That provides a new, combinatorial interpretation of the degree.

We have taken this as an opportunity to review the most recent results on tensor models applied in the context of the loop models. This approach also unravels the challenges faced by random tensor theory. It is emphasized that to our knowledge there is no known solution to those models (e.g. for the large $N$ free energy at finite $\tau$), beyond a very particular case which corresponds to a Penner model. Beyond this case, the most generic and explicit result is the classification of edge-colored graphs according to their degree, due to Gurau and Schaeffer \cite{GurauSchaeffer} which as we have explained in section \ref{sec:ScalingLimits} classifies the loop configurations at fixed genus and number of edges according to the number of loops. There is moreover a double-scaling limit which sums consistenly the most singular (at criticality) loop configurations. We hope that the relationship we have established between tensor models and loop models can lead to fruitful cross-fertilization.

While we have focused in section \ref{sec:ScalingLimits} on the scaling limits, further connections between matrix and tensor models have been reviewed in section \ref{sec:Bimaps}, based on the Hubbard-Stratanovich (intermediate field) transformation. It reveals that melonic quartic tensor models generate maps formed by maps with different edge colors glued together at vertices (or by duality, at the center of their faces), \cite{DSQuartic} (see also \cite{BeyondPert} for a constructive analysis (Borel summability) of this model and \cite{GenericQuartic} for an extension of those ideas to arbitrary quartic models). In the case of two edge colors, those maps have already appeared under the name of \emph{nodal surfaces} in \cite{EynardBookMulticut} as multicut solutions of the one-matrix model\footnote{Obviously the multicut solution satisfies the loop equations of the 1-matrix model and is not a solution of quartic tensor models which correspond to a different evaluation of the generating function of nodal surfaces.}. It has been further observed that such maps can be generated in a Givental-like fashion\footnote{We are indebted to Bertrand Eynard for pointing this out and we would like to thank St\'ephane Dartois for sharing his progress on such a re-formulation of tensor models.} \cite{GiventalDartois}.

We believe that viewing tensor models as matrix models constitutes an interesting research road and places tensor models in a frame where powerful tools are available. In particular, the intermediate field method turns quartic tensor models into matrix models which generate generalizations of nodal surfaces. Either techniques developed for matrix models, such as the topological recursion \cite{TopRec}, or combinatorial approaches, could lead to new results. Among the combinatorial approaches, bijective methods akin to Schaeffer's bijection for planar quadrangulations could be useful to solve the large $N$ limit (i.e. the planar sector) of the two-matrix model \eqref{TwoMatrixModel}, while algebraic methods have proved helpful to probe maps at arbitrary genus \cite{KPGouldenJackson, CarrellChapuyRecursion}. Preliminary calculations suggest that the large $N$ limit of \eqref{TwoMatrixModel} is a generalization of the $O(\tau)$ model (where the eigenvalues of $M_1$ are attracted to the mirror image of those of $M_2$) \cite{ExactO(n)Eynard, O(n>2)Eynard}.

\section*{Acknowledgements}

Research at Perimeter Institute is supported by the Government of Canada through Industry Canada and by the Province of Ontario through the Ministry of Research and Innovation.

\appendix

\section{Interpolating $1/N$ expansions in tensor models} \label{sec:interpolating}

We have shown in section \ref{sec:loops} that the $U(\tau)$ loop models are tensor models in disguise whose ordinary scaling with $N$ is recovered when the number of matrices scales as the size of the matrices. Yet, other choices of scalings might make sense and the question is then whether they bring new behaviors or not. To answer this question, we show in \ref{sec:d=3Interpolation} that it is indeed possible to set $\tau = N^\beta$ with $\beta\in[0,1]$, or in other words to work with a random tensor of size $N\times N\times N^\beta$.

We will then generalize this approach to tensors of higher ranks, equipped with the standard scaling of tensor models in section \ref{sec:TensorBetaScaling}, and with the new scaling introduced in \cite{new1/N} in section \ref{sec:OtherScalings}. Generally, some tensor indices have range $N$ and others have range $N^\beta$, or they all have range $N$ but the powers of $N$ in the action are $\beta$-dependent. In all cases it is found that as soon as $\beta >0$ the behavior is the same as for $\beta=1$ and only $\beta=0$ is different. However, the amplitude of the Feynman graphs depend on $\beta$-dependent linear combinations of degrees and/or genera of subgraphs. As a consequence, the order in which the $1/N$ corrections appear depends on $\beta$.

\subsection{Intermediate scalings in the loop models} \label{sec:d=3Interpolation}

The scaling in front of the action has to be $N^{1+\beta}$ instead of $N$ in \eqref{matrixF}. The loop fugacity being $N^{\beta}$, a Feynman graph receives a factor $N^{\beta L}$. Therefore the exponent of $N$ in a Feynman graph is
\be
F+\beta L - (1+\beta)(E-V).
\ee
This leads to the free energy
\begin{equation}
	N^{2+\beta} f = \sum_{\substack{\text{connected}\\ \text{$4$-colored graphs}}} N^{\beta(3-\omega(G))+(1-\beta)(2-2g(G))}\ \tilde{\lambda}^{E-V}\ \frac1{s(G)}.
\end{equation}
As expected it scales extensively, with the number of degrees of freedom $N^{2+\beta}$, so that $f$ is finite at large $N$. For $\beta=0$, the standard scaling of matrix models is reproduced. As soon as $\beta>0$, since both quantities $\beta\omega(G)$ and $(1-\beta)g(G)$ are positive, the large $N$ limit projects onto graphs such that $\omega(G) = g(G) = 0$. This defines the same large $N$ limit as for $\beta=1$, dominated by the CPL configurations which are melonic.

\begin{figure}
		\subfigure[Graph $G_1$ with $\omega(G_1) = 4$ and $g(G_1) = 1$]{\includegraphics[width=4cm]{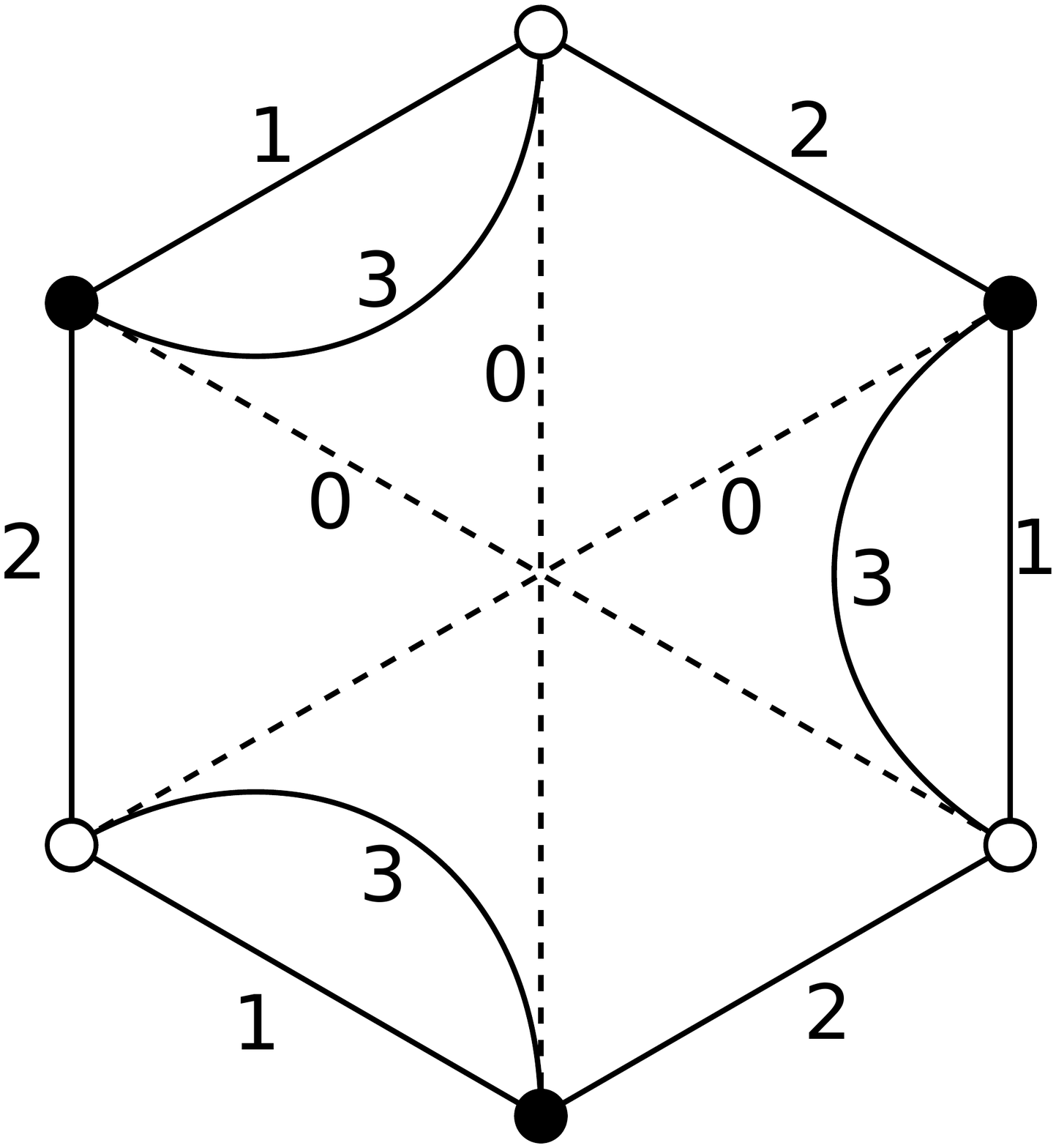}}
	\hspace{1cm}
		\subfigure[Graph $G_2$ with $\omega(G_2) = 5$ and $g(G_2) = 0$]{\includegraphics[width=4cm]{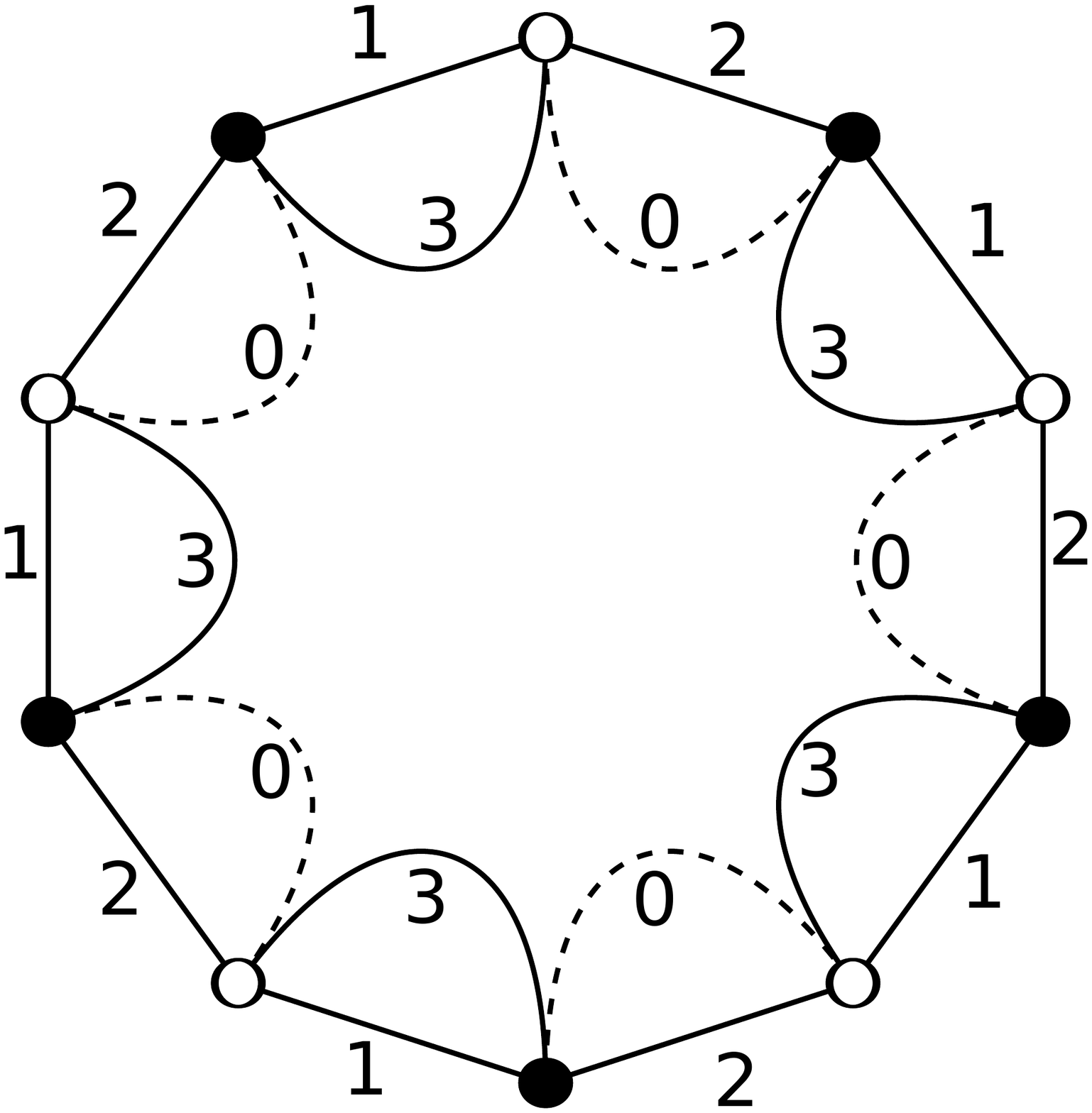}}
	\caption{The graph $G_1$ (on the left) dominates the graph $G_2$ (on the right) as long as $\beta < \frac{1}{3}$. They contribute to the same order when $\beta = \frac{1}{3}$, then for higher values of $\beta$, $G_2$ dominates. \label{fig:BetaDependence}}
\end{figure}

Yet, while the leading order at large $N$ is the same for all $\beta > 0$, the higher orders in the $1/N$ expansion depend on the value of $\beta$: if two graphs $G_1$ and $G_2$ are such that $g(G_1) < g(G_2)$ and $\omega(G_1) > \omega(G_2)$, then, when $\beta = 0$, $G_1$ contributes to a lower order than $G_2$, as it has a lower genus, but when $\beta$ reaches $1$, the contribution of $G_2$ dominates, for the degree of $G_1$ is higher. Between these two situations, there is a value of $\beta$,
\begin{equation}
	0 < \beta = \frac{\omega(G_1) - \omega(G_2)}{2(g(G_2) - g(G_1)) + \omega(G_1) - \omega(G_2)} < 1,
\end{equation}
such that both graphs contribute at the same order. Such graphs do exist, an example is given in figure \ref{fig:BetaDependence}.

Thus, if intermediate scalings with $\beta > 0$ do not show any $\beta$-dependence at leading order, higher orders do depend on $\beta$.

The single trace invariants of the matrix action correspond to polynomials associated to 3-colored bubbles with only one face of colors $(1,2)$. It is also possible to introduce bubbles with more faces of colors $(1,2)$, provided they are appropriately scaled in the action according to their number of faces (similarly, to define a matrix action with multi-trace invariants, i.e. disconnected loops, these terms must be re-scaled by $1/N$ to the number of traces).

\subsection{Tensor models for tensors of size $N\times \dotsm \times N\times N^\beta$} \label{sec:TensorBetaScaling}

\subsubsection{The standard scaling} \label{sec:standardscaling}

Bubbles and colored graphs are the ingredients to build random tensor models. To each bubble, an invariant polynomial in the tensor entries can be built, and the Feynman graphs of tensor models are precisely colored graphs built from the bubbles. We provide here a brief summary of their construction.

Let $I$ be a finite set, and $\{B_i\}_{i\in I}$ be a set of bubbles. We denote $B_i(T,\overline{T})$ the corresponding invariant polynomials, for $T$ a tensor of size $N^d$. The tensor action is
\begin{equation} \label{tensoraction}
	S(T, \overline{T}) = T \cdot \overline{T} + \sum_{i\in I} t_i\ B_i(T, \overline{T}).
\end{equation}
where $T \cdot \overline{T} = \sum T_{i_1\dotsb i_d}\overline{T}_{i_1\dotsb i_d}$ is the quadratic part. The partition function $Z$ and the free energy $f$ are
\begin{equation}
\label{tensor F}
	Z = e^{-f} = \int \diff T \diff \overline{T} \exp\left(-\frac{N^{d-1}}{\lambda}S(T, \overline{T})\right).
\end{equation}
The free energy admits the expansion onto connected graphs $G$. The Feynman rules require to connect the vertices of bubbles (which carry $T$s and $\overline{T}$s) with lines corresponding to the bare covariance. Giving these lines the fictitious color 0, the connected Feynman graphs are precisely $(d+1)$-colored graphs. Such a graph $G$ is made of $b_i$ bubbles of type $i\in I$. Its total number of bubbles is $b=\sum_{i\in I} b_i$, its number of vertices is $2p$, and it contains $F_{0a}$ faces with colors $(0,a)$. The $N$-dependence of the amplitude of such a graph comes with a factor $N^{d-1}$ per bubble, a factor $N^{-(d-1)}$ per line of color $0$ (there are $p$ of them), and there is a free sum per face of colors $(0,a)$ which brings a factor $N$. Thus the exponent of $N$ in the amplitude of $G$ is $\sum_a F_{0a} - (d-1)(p-b)$. Using the formula \eqref{graph degree}, the following expansion holds
\begin{equation}
	f = \sum_{\substack{\text{connected} \\ \text{$(d+1)$-colored graphs $G$}}} N^{d-\frac{2}{(d-1)!} \omega(G) + \frac{2}{(d-2)!}\sum_{i\in I} b_i \omega(B_i)}\ \frac{1}{s(G)}\ \lambda^{p-b}\ \prod_{i\in I} (-t_i)^{b_i}.
\end{equation}
As explained in \cite{uncoloring}, a $\Delta$-colored graph is dual to a triangulation of a $(\Delta-1)$-dimensional pseudo-manifold. This is done by assigning a simplex of dimension $\Delta-1$ to each vertex, and for any line which connects two vertices, we glue the corresponding simplices along some of their faces\footnote{The gluing is unambiguous thanks to the coloring \cite{uncoloring}.} Therefore the bubbles used in the action \eqref{tensoraction} are dual to simplices of dimension $d-1$. By taking the topological cone over the dual triangulation to a bubble, one creates a \emph{chunk} of space in dimension $d$. As $(d+1)$-colored graphs, the Feynman graphs of the expansion of the free energy are dual to triangulations of $d$-dimensional pseudo-manifolds. They are obtained from the chunks dual to the bubbles by gluing them along some faces (which correspond to the lines of color 0).

\subsubsection{Interpolating scaling}

Suppose that we have a tensor model which generates $(d+1)$-colored graphs, with bubbles which have a single connected component of colors $1,\dotsc,d-1$. In the expansion of the free energy, there is a single subgraph with colors $0,1,\dotsc,d-1$ for each Feynman graph. These sub-graphs are dual to triangulations of dimension $d-1$. The last index of the tensor, in position $d$, creates faces with colors $0,d$. If it has a range $a_d=1,\dotsc,\tau$, we can interpret those faces as loops on $d$-colored graphs (with colors $0,1\dotsc,d-1$). The color $d$ of each bubble corresponds to a portion of a loop which goes through the chunk dual to the bubble. Loops are then obtained when the chunks are glued. Therefore the triangulations of dimension $d$ generated by such a tensor model can be seen as triangulations in dimension $d-1$ decorated with loops. This is the extension of the correspondence exhibited in section \ref{sec:loops}.

Let $G$ be a $(d+1)$-colored Feynman graph with $2p$ vertices, $b_i$ bubbles $B_i$, $i\in I$ and $b=\sum_{i\in I}b_i$ the total number of bubbles in $G$. Its degree $\omega_d(G)$ counts the total number of faces with colors $(0,a)$ for all $a=1,\dots,d$. The degree $\omega_{d-1}(G)$ of the $d$-colored subgraph with colors $0,\dotsc,d-1$ counts the number of faces with colors $(0,a)$ for $a=1,\dotsc,d-1$. Therefore, applying \eqref{graph degree} to these two graphs, we can extract the number of faces $F_{0d}$, i.e. the number of loops on the triangulation dual to the subgraph with colors $0,1,\dots,d-1$,
\be
F_{0d} = p-b +1 -\frac2{(d-1)!}\omega_d(G) + \frac2{(d-2)!}\sum_{i\in I} b_i \omega_d(B_i) + \frac2{(d-2)!} \omega_{d-1}(G) - \frac2{(d-3)!} \sum_{i\in I} b_i\omega_{d-1}(B_i).
\ee
This is the generalization to arbitrary $d$ of the counting of loops \eqref{eq:LoopCounting} at $d=3$. Here $\omega_d(B_i)$ is the degree of the bubble $B_i$ and $\omega_{d-1}(B_i)$ the degree of its sub-bubble with colors $1,\dotsc,d-1$.

We consider a tensor with components $T_{a_1\dotsb a_d}$ with $a_j=1,\dotsc,N$ for $j=1,\dotsc,d-1$ and $a_d$ of range $\tau N^\beta$ for $\beta\in[0,1]$. We set the scaling in front of the action to $N^{d-2+\beta}$ instead of $N^{d-1}$ in \eqref{tensor F}. For each Feynman graph $G$, the exponent of $N$ is
\begin{multline}
\sum_{i=1}^{d-1} F_{0i} + \beta F_{0d} - (d+\beta-2)(p-b) \\
= \beta\left(d-2\frac{\omega_d(G)}{(d-1)!} + 2 \sum_{i\in I}\frac{b_i \omega_d(B_i)}{(d-2)!}\right) + (1-\beta)\left(d-1 - 2\frac{\omega_{d-1}(G)}{(d-2)!} + 2 \sum_{i\in I}\frac{b_i \omega_{d-1}(B_i)}{(d-3)!}\right).
\end{multline}
For $\beta=0$, we obviously recover the scaling of the rank $d-1$ tensor model, dominated at large $N$ by graphs which are melonic on the colors $0,1,\dotsc,d-1$, $\omega_{d-1}(G) = \omega_{d-1}(B_i)=0$. In particular, the bubbles need not be melonic on all the colors in the sense that the lines of colors $d$ can be placed in any possible way in the bubbles $\{B_i\}$. The only dependence of the amplitude on them is through $\tau^{F_{0d}}$. But as soon as $\beta>0$, the leading order requires $\omega_d(G) = \omega_d(B_i)=0$ too, which means that this restricts to the melonic $(d+1)$-colored graphs.

Nevertheless, the higher orders may depend on $\beta$, according to the balance between the degree of the subgraphs of colors $(0,1,\dotsc,d-1)$ and $(0,1,\dotsc,d)$, just like in the case $d=3$ of section \ref{sec:d=3Interpolation}. To understand at which level $\beta$ plays a role, the classification of edge-colored graphs can be applied. Clearly, the melonic 2-point subgraphs on $d+1$ colors are also melonic on the colors $(0,\dotsc,d-1)$. The chains on $d+1$ colors can be either chains or melonic subgraphs on the colors $(0,\dotsc,d-1)$. Therefore the reduction used in \cite{GurauSchaeffer} from colored graphs to schemes via melon removals and chain contractions is unchanged (and the singularities of the generating function is still controlled by the number of broken chains on the colors $(0,\dotsc,d+1)$). This means that the way varying $\beta$ shuffles the $1/N$ expansion has to be investigated at the level of the schemes themselves, which requires an analysis beyond the present article.

\subsection{Other scalings of tensor models} \label{sec:OtherScalings}

\subsubsection{Probing sub-graphs}

Other scalings have been proved to lead to a well-defined large $N$ limit \cite{new1/N}. The idea is to probe the colored graphs not in terms of their degree, but in terms of the degrees of subgraphs carrying different subsets of colors. For instance, for $d\geq 5$, we can split the set of colors $\{1,\dotsc,d\}$ into two subsets with at least two colors, $D_1=\{1,\dotsc,d_1\}$ and $D_2 = \{d_1+1,\dotsc,d\}$, with $2\leq d_1\leq d-2$. We can relate the degree of a $(d+1)$-colored graph $G$ to the degree of the subgraph $G_{D_1}$ with colors $0,1,\dotsc,d_1$ and the degree of the subgraph $G_{D_2}$ with colors $0,d_1+1,\dotsc,d$,
\begin{equation}
\label{eq:RelatingDegrees}
\begin{aligned}
	& d - 2\left(\frac{1}{(d-1)!} \omega(G) - \frac{1}{(d-2)!}\sum_{i\in I} b_i \omega(B_i)\right) \\
	& = \sum_{a=1}^d F_{0a} - (d-1)(p-b) = \sum_{a=1}^{d_1} F_{0a} - (d_1-1)(p-b) + \sum_{a=d_1+1}^d F_{0a} - (d-d_1-1)(p-b) -(p-b) \\
	& = d - 2\left(\frac{\omega(G_{D_1})}{(d_1-1)!} - \sum_{i\in I} \frac{b_i\omega(B_{i,D_1})}{(d_1-2)!}\right) - 2\left( \frac{\omega(G_{D_2})}{(d-d_1-1)!} - \sum_{i\in I} \frac{b_i\omega(B_{i,D_2})}{(d-d_1-2)!}\right) - (p-b).
\end{aligned}
\end{equation}
Here $B_{i, D_1}, B_{i, D_2}$ are the sub-bubbles of the bubble $B_i$ with colors in $D_1, D_2$. Therefore the difference between scaling with the degree of $G$ and scaling with the degrees of the subgraphs is a term $N^{-(p-b)}$. Since the amplitude of graphs also displays a term $\lambda^{p-b}$, this suggests to scale $\lambda$ like $N$, $\lambda = N \widetilde{\lambda}$ with $\widetilde{\lambda}$ finite. Consequently, the scaling in front of the action in \eqref{tensor F} becomes $N^{d-1}/\lambda = N^{d-2}/\widetilde{\lambda}$.

This provides the intuition of the new $1/N$ expansion presented in \cite{new1/N}. To be precise, it is however necessary to be more careful due to the fact that sub-bubbles and subgraphs might have several connected components while the bubbles and the graphs themselves are connected. This forces to re-scale some bubbles in the action to avoid unboundedness of the free energy,
\be
\int [dT\,d\overline{T}]\ \exp -\frac{N^{d-2}}{\widetilde{\lambda}} \left( T\cdot \overline{T} + \sum_{i\in I} N^{2-n(B_{i,D_1})- n(B_{i, D_2})}\,t_i\,B_i(T,\overline{T}) \right),
\ee
where $n(B_{i,D_{1,2}})$ is the number of connected components of the subgraphs with colors in $D_{1,2}$ of $B_i$.

This process can be repeated to probe more than two types of subgraphs, as long as the corresponding subsets of colors contains at least two colors.

This approach in fact enables to define tensor models for `rectangular' tensors, of size $N_1^{D_1}\times \dotsm \times N_L^{D_L}$, where $\sum_{l=1}^L D_l = d$ and $D_l \geq2$ is the number of indices with range $N_l$, \cite{new1/N}.

However a question left unanswered in \cite{new1/N} is what happens for a tensor which has a single index which scales independently of all the others. Indeed, if an index, say in position $k$, has range $\tau$, it creates in the Feynman graphs a factor $\tau^{F_{0k}}$ and the number of faces $F_{0k}$ cannot be packed into a genus or a degree. We have actually solved this problem in the previous sections, interpreting the $(d+1)$-colored graphs as triangulations in dimension $d-1$ decorated with loops.



\subsubsection{From random matrices to random tensors of size $N^2 \times N^{2\beta}$}

The above scalings define tensor models with `slices' of colors, each slice having a parameter $N_i$. In \cite{new1/N}, the case where they all scale together at the same rate $N$ was emphasized. Here we investigate the case where they do not have the same rate with $N$. For instance, we can interpolate between $d = 2$ (matrix model, or 2-colored bubbles)  and $d = 4$ (4-colored bubbles) by taking a tensor of rank 4 and using the standard scaling on the matrix part (i.e. the colors (1,2)) and a scaling $N^\beta$ on the colors (3,4), with $\beta\in[0,1]$.

We consider 4-colored bubbles which have only one connected components on the colors 1,2 and on the colors 3,4 (we can make sense of the model for arbitrary bubbles if we scale them in the action with a factor $N^{1+\beta-F_{12}-\beta F_{34}}$, using techniques developed in \cite{new1/N}). The scaling in front of the action has to be $N^{1+\beta}$. This way, the exponent of $N$ for a $(4+1)$-colored graph $G$ in the Feynman expansion of the free energy is
\begin{equation}
	F_{01}+F_{02} + \beta(F_{03}+F_{04}) - (1+\beta)(p-b) = \left(2-2g_{012}(G)\right) + \beta\left(2 - 2g_{034}(G)\right)
\end{equation}
For $\beta=0$, this is obviously a standard one-matrix model, dominated by planar graphs at large $N$. As soon as $\beta > 0$, the leading order graphs are those whose subgraphs with colors 0,1,2 and with colors 0,3,4 both have vanishing genus, just like in the case $\beta=1$. But once again, the higher orders do depend on the actual value of $\beta$.

\subsubsection{Interpolating scaling at fixed tensor size}

The $1/N$ expansions for rectangular tensors introduced in \cite{new1/N} are all well-defined in the particular case of `square' tensors, when $a_i=1,\dotsc,N$ for all indices. Therefore it should be possible to interpolate them, and investigate the intermediate regimes.

For a rank 4 tensor of size $N\times N\times N\times N$, we have at our disposal the standard scaling summed up in section \eqref{sec:standardscaling} (with a factor $N^3$ in front of the action), but also a scaling based on two color slices $D_1=\{1,2\}, D_2=\{3,4\}$ for which the Feynman graphs scale with the genera of the sub-graphs with colors $0,1,2$ and those with colors $0,3,4$ (the factor in front of the action is $N^2$).

The $\beta$-dependent free energy is defined by
\be
f_\beta = -\ln \int [dT\,d\overline{T}]\ \exp -N^{2+\beta} \left(T\cdot\overline{T} + \sum_{i\in I} N^{\beta(2-n(B_{i,D_1}) -n(B_{i,D_2}))}\,t_i\,B_i(T,\overline{T}) \right).
\ee
To keep things simple, we are going to assume that the action is a superposition of bubble polynomials $\{B_i\}_{i\in I}$ for bubbles which have a single face with colors $(1,2)$ and a single face with colors $(3,4)$ (i.e. a single connected component on the colors 1,2, $n(B_{i,D_1})=1$ and on the colors 3,4, $n(B_{i,D_2})=1$).

The Feynman expansion generates $(4+1)$-colored graphs. Each face of colors $(0,a)$, $a=1,2,3,4$, brings a factor $N$, while an insertion of the bubble $B_i$ brings $N^{2+\beta}$ and a line of color 0 ($p$ of them) gives $N^{-(2+\beta)}$. By writing the total number of faces as $\beta \sum_a F_{0a} + (1-\beta)\sum_a F_{0a}$, and writing $2+\beta = 3\beta+ 2(1-\beta)$, we see that the exponent of $N$ in a (connected) Feynman graph reads
\begin{equation}
\begin{aligned}
  \sum_{i=1}^4 F_{0i} - \left(2+\beta\right)(p-b) &= \beta \left(\sum_{i=1}^4 F_{0i} - 3(p-b)\right) + (1-\beta) \left( \sum_{i=1}^4 F_{0i} - 2(p-b)\right),\\
  &=\beta\left(4 - 2\frac{\omega(G)}{(4-1)!} + 2\sum_{i\in I}\frac{b_i \omega(B_i)}{(4-2)!} \right) + (1-\beta)\bigl(2 - 2g_{012}(G) + 2 - 2g_{034}(G)\bigl).
\end{aligned}
\end{equation}
Here $g_{012}$ (resp. $g_{034}$) is the degree of the sub-graph $G_{D_1}$ with colors $(0,1,2)$ (resp. $G_{D_2}$ with colors $(0,3,4)$). For $\beta=0$ this coincides with the new $1/N$ expansion proposed in \cite{new1/N}, and for $\beta=1$ with the standard rank 4 tensor model scaling. For any $\beta>0$, the leading order contributions are graphs with vanishing degree $\omega(G)=0$. Therefore only the case $\beta=0$ is not dominated by melonic graphs only (but by the larger set of graphs which are planar on $0,1,2$ and planar on $0,3,4$). As usual, the higher orders do depend on $\beta$.

\end{document}